\documentclass[aip,jcp,reprint,amssymb]{revtex4-1}
\usepackage{graphicx}
\usepackage[caption=false]{subfig}
\usepackage{xcolor}
\newlength{\myfigwidth}
\usepackage{amsmath}
\begin{document}
\title{Role of Non-Equilibrium Conformations on Driven Polymer Translocation}
\author{H. H. \surname{Katkar}}
\altaffiliation{Present address: Department of Chemistry, The University of Chicago, Chicago, IL 60637, USA}
\affiliation{Department of Polymer Science and Engineering, University of Massachusetts, Amherst, MA 01003, USA}
\author{M. Muthukumar}
\email{muthu@polysci.umass.edu}
\affiliation{Department of Polymer Science and Engineering, University of Massachusetts, Amherst, MA 01003, USA}
\begin{abstract}
One of the major theoretical methods in understanding polymer translocation through a nanopore is the Fokker-Planck formalism based on the assumption of quasi-equilibrium of polymer conformations. The criterion for applicability of the quasi-equilibrium approximation for polymer translocation is that the average translocation time per Kuhn segment, $\langle \tau \rangle/N_K$ is longer than the relaxation time $\tau_0$ of the polymer. Towards an understanding of conditions that would satisfy this criterion, we have performed coarse-grained three dimensional Langevin dynamics and multi-particle collision dynamics simulations. We have studied the role of initial conformations of a polyelectrolyte chain (which were artificially generated with a flow field) on the kinetics of its translocation across a nanopore under the action of an externally applied transmembrane voltage $V$ (in the absence of the initial flow field). Stretched (out-of-equilibrium) polyelectrolyte chain conformations are deliberately and systematically generated and used as initial conformations in translocation simulations. Independent simulations are performed to study the relaxation behavior of these stretched chains and a comparison is made between the relaxation timescale and the mean translocation time ($\langle \tau \rangle$). For such artificially stretched initial states, $\langle \tau \rangle/N_K < \tau_0$, demonstrating the inapplicability of the quasi-equilibrium approximation. Nevertheless, we observe a scaling of $\langle \tau \rangle \sim 1/V$ over the entire range of chain stretching studied, in agreement with the predictions of the Fokker-Planck model. On the other hand, for realistic situations where initial artificially imposed flow field is absent, a comparison of experimental data reported in the literature with  the theory of polyelectrolyte dynamics reveals that the Zimm relaxation time ($\tau_\text{Zimm}$) is shorter than the mean translocation time for several polymers including ssDNA, dsDNA and synthetic polymers. Even when these data are rescaled assuming a constant effective velocity of translocation, it is found that for flexible (ssDNA and synthetic) polymers with $N_\text{K}$ Kuhn segments, the condition $\langle\tau\rangle/N_\text{K} < \tau_\text{Zimm}$  is satisfied. We predict that for flexible polymers such as ssDNA, a cross-over from quasi-equilibrium to non-equilibrium behavior would occur at $N_\text{K} \sim O(1000)$.

\end{abstract}
\maketitle
\section{Introduction}
\begin{figure}
	\centering
		\includegraphics*[width=1.25\myfigwidth]{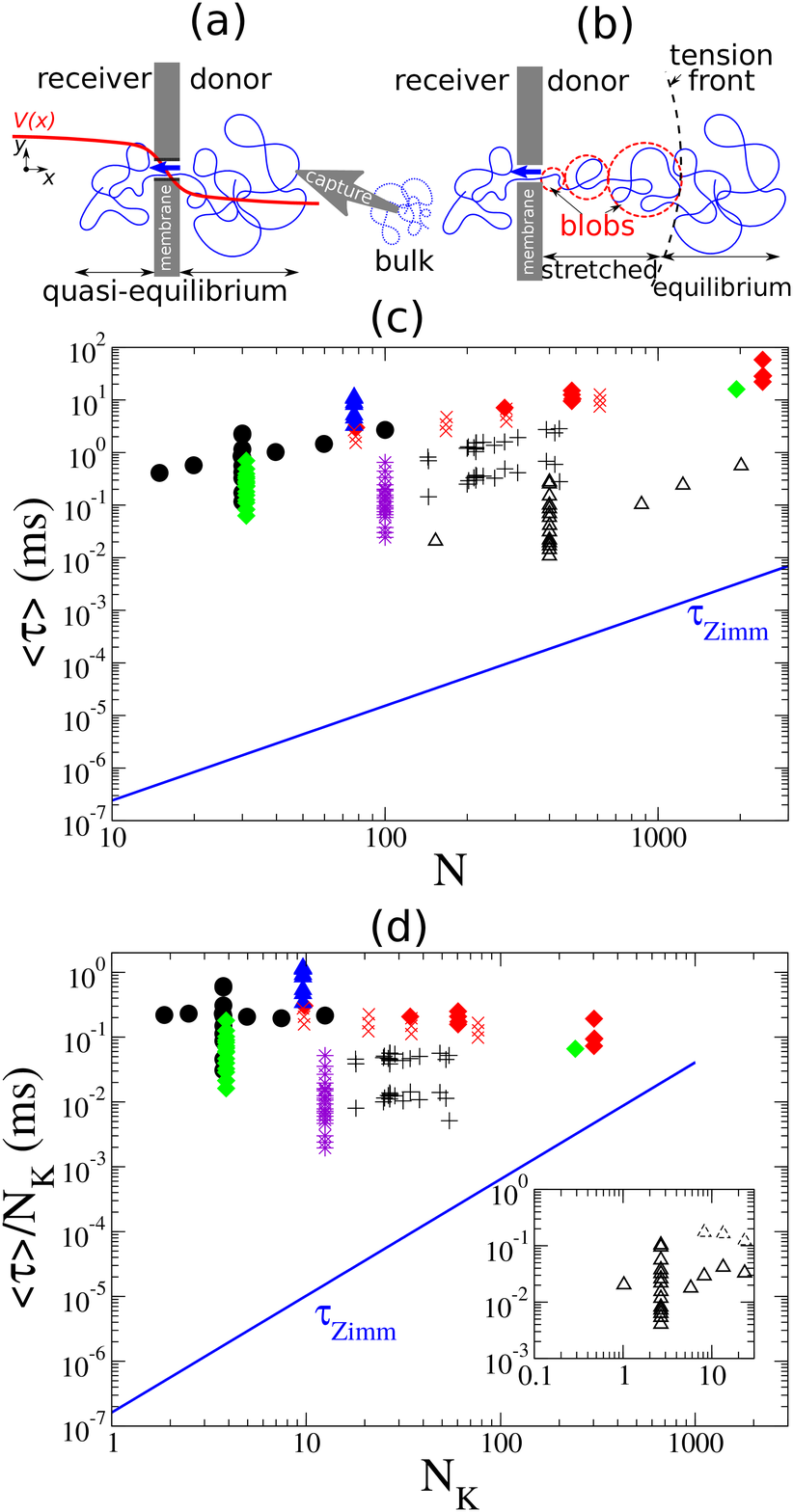}
		\caption{(a) Schematic for the translocation setup showing a negatively charged polyelectrolyte chain (blue) captured and being translocated through the nanopore (in the -ve $x$-direction). The red curve shows a typical voltage profile $V(x)$. The Fokker-Planck model assumes that the chain is in quasi-equilibrium during the entire translocation. (b) Schematic for the tension propagation model showing the stretched part of the chain on the donor side that forms frictional blobs of increasing size. (c) Comparison of Zimm relaxation time $\tau_\text{Zimm}$ (blue line, calculated using a 2 nm Kuhn length and a monomer size of {$2.5$ \AA }) with experimentally measured mean translocation time $\langle\tau\rangle$ (black plus\cite{Kasianowicz1996}, blue triangle \cite{Wong2010}, red cross \cite{Jeon2014a}, green diamond \cite{Brun2008}, red diamond \cite{Murphy2007}, black circle \cite{Meller2001}, violet asterisk \cite{Meller2000} and open black triangle \cite{Wanunu2008}) as a function of chain length $N$ (see text for details). A clear separation of translocation and relaxation timescales is evident for polymers of length up to $N\sim3000$. (d) Comparison of $\tau_\text{Zimm}$ (blue line) with rescaled translocation time from (c). The experimental translocation time is divided by the number of Kuhn segments $N_K$. Inset shows similar plot for a dsDNA translocating through a solid-state nanopore , assuming $150$ base-pairs per Kuhn segment instead. Note that in Ref. \cite{Wanunu2008}, the translocation event data obtained in a single experiment is divided into faster (triangles) and slower (open triangles, not included in (c)) translocating populations.}
		\label{fig:fig1}
\end{figure}

Voltage driven single-file translocation of a polyelectrolyte across nanopores has received broad attention due to its potential application as a high-throughput cost-effective alternative for DNA sequencing \cite{Kasianowicz1996} and due to the rich physics involved \cite{Muthukumar2011}. A typical experimental setup consists of two reservoirs containing a salt solution, separated by a membrane embedded with a nanopore. Polyelectrolyte added to the donor reservoir translocates across the nanopore due to an applied transmembrane voltage $V$ (Figure \ref{fig:fig1}(a)). The capture of the polyelectrolyte from the bulk involves an entropic barrier \cite{Muthukumar2011} that can be overcome by an extended weak electric field \cite{Wanunu2010}, assisted by electro-osmotic flow \cite{Wong2010,Rowghanian2013,Rowghanian2013a} and electrostatic attraction from the nanopore \cite{Jeon2014}. Once the translocation process is nucleated, there is a finite probability that the chain translocates successfully across the nanopore instead of being rejected back to the donor reservoir. The translocation process is stochastic due to coupled counter-ion dynamics and binding kinetics with the nanopore \cite{Cohen2012,Katkar2014,Banerjee2014}, in addition to thermal fluctuations. Moreover, the chain can have different conformations when captured \cite{Izmitli2008,Saito2012,Sarabadani2014,Farahpour2013,Vollmer2016}, contributing towards the stochastic nature of the process.
Translocation of polyelectrolytes is a non-equilibrium phenomenon due to the externally applied electric field.  There have been two distinct approaches towards modeling this phenomenon, based on the relative magnitudes of the average time for translocation and the conformational relaxation time for a polyelectrolyte chain.
If the polyelectrolyte chain is relaxed at the timescale at which it translocates, a quasi-equilibrium assumption can be made. Various approaches have been suggested to increase the translocation time relative to the relaxation time and thereby ensuring quasi-equilibrium, such as changing experimental parameters like solvent viscosity\cite{Fologea2005,deHaan2012}, nanopore-polymer interactions \cite{Ding2012,Rasmussen2012,Maglia2008,Anderson2013,Banerjee2014}, and salt identity\cite{Kowalczyk2012} among several others.  If, on the other hand, the chain relaxation is not fast enough, non-equilibrium effects have to be taken into consideration. Additionally, the chain conformations just after capture can be far from equilibrium at the onset of translocation because of the strong electro-osmotic flow (predicted \cite{Wong2007} to be as high as $35$ cm/s for a typical charge density of $0.14$ e/nm$^2$ on the pore surface) and the extended electric field near the pore mouth. 

We now briefly summarize the experimental results from the literature on the average translocation time for a variety of flexible polyelectrolyte chains. On the other hand, the relaxation time for a polyelectrolyte chain is yet to be experimentally established. However, the Zimm relaxation time, generalized from the original expression for Gaussian chains to polyelectrolyte chains, is a reasonable measure of the chain relaxation time. The experimentally measured mean translocation time $\langle\tau\rangle$ is shown in Figure \ref{fig:fig1}(c) as a function of the number of monomers per chain $N$, from translocation experiments in the literature \cite{Kasianowicz1996, Jeon2014a, Brun2008, Wong2010, Murphy2007, Meller2001, Meller2000}. An $\alpha$-Hemolysin ($\alpha$-HL) nanopore is used in all these experiments (except in Ref. \cite{Wanunu2008}, where a SiN nanopore is used), with the polyelectrolyte being NaPSS \cite{Wong2010,Murphy2007,Jeon2014a}, dextran sulfate sodium \cite{Brun2008}, ss-DNA \cite{Kasianowicz1996,Meller2000,Meller2001} or dsDNA \cite{Wanunu2008}.  Multiple data points for a given $N$ correspond to variation in other parameters (voltage \cite{Brun2008,Jeon2014a,Kasianowicz1996,Meller2001,Murphy2007,Wanunu2008}, temperature \cite{Meller2000} and pH \cite{Wong2010}). The applied voltage $V$ across the $\alpha$-HL nanopore in these experiments is $40$-$260$ mV, while for the $10$ nm long solid-state nanopore \cite{Wanunu2008}, $V \equiv 200$-$500$ mV. 

The Zimm relaxation time for a Gaussian chain with $N_K$ Kuhn segments, each of length $\ell$, is given by \cite{Doi1987}.
\begin{equation}
\tau_\text{Zimm} = \frac {\eta \ell^3}{\sqrt {3 \pi} k_BT} N_K^{3/2} = 6 \sqrt {\frac {2}{\pi}} \frac {\eta}{k_BT} R_{g0}^3,
\end{equation}
where $k_BT$ is the Boltzmann constant times the absolute temperature, $\eta$ is the viscosity of the solution, and $R_{g0}$ is the radius of gyration of a Gaussian chain given by $R_{g0}=\sqrt{N_K \ell^2/6}$. We assume that the above equation is valid also for a polyelectrolyte chain by replacing $R_{g0}$ by $R_g$, 
\begin{equation}
\tau_\text{Zimm} = 6 \sqrt {\frac {2}{\pi}} \frac {\eta}{k_BT} R_{g}^3,
\end{equation}
where $R_g$ is the radius of gyration of a polyelectrolyte chain. In general,
\begin{equation}
R_g = \ell f(\chi, \kappa) N_K^\nu,
\end{equation}
where $f(\chi, \kappa)$ is a function of the Flory-Huggins excluded volume parameter $\chi$, the Debye length $\kappa^{-1}$, and other characteristics of the charged polymer and the electrolyte solution. The value of the size exponent $\nu$ depends on the level of electrostatic screening in the system. For almost all experiments on single molecule electrophoresis through nanopores, the electrolyte concentration in the solution is quite high ($\sim$1 M monovalent salt). For such conditions, $\nu =3/5$, and $R_g$ is given by \cite{Muthukumar1987,Beer1997,Muthukumar2011}
\begin{equation}
\frac {R_g}{\ell} \simeq 0.34 [(\frac {1}{2} - \chi) + \frac {4\pi \alpha_p^2 z_p^2 \ell_B}{\kappa^2 \ell^3}]^{1/5} N_K^{3/5}.
\end{equation}
Here $\alpha_p$ is the average degree of ionization of the chain, $z_p$ is the net charge per Kuhn segment, $\ell_B$ is the Bjerrum length ($=e^2/(4\pi \epsilon_0 \epsilon k_BT)$, where $e$ is the electronic charge, $\epsilon_0$ is the permittivity of vacuum, and $\epsilon$ is the dielectric constant of the solution). $N_K$ is proportional to the number of monomers $N$ in the chain and $\ell$ is proportional to the monomer size $b$ \cite{Flory1969,Yamakawa1971}. By taking $\chi=1/2, \alpha_p =0.2, z_p=8, \ell_B = 0.7$ nm, $\kappa^{-1}=0.304$ nm, $\ell = 2$ nm, and $N_K = N/8$, typical of ssDNA or NaPSS \cite{Nishida1997,Chi2013} in 1M monovalent salt solution,
\begin{equation}
R_g \simeq 0.6  N^{3/5} b.
\end{equation}
Substituting Eq. (5) into Eq. (2), we get
\begin{equation}
\tau_\text{Zimm} \simeq 1.03 \frac {\eta b^3}{k_BT} N^{9/5}.
\label{eqn:Zimm}
\end{equation}
Using $T=300$ K, $\eta = 0.001$ Pa$\cdot$s, and $b=2.5$ \AA\text{ }, the Zimm relaxation time for the range of $N$ is shown by a solid blue line in Figures \ref{fig:fig1}(c) and \ref{fig:fig1}(d) .

As evident from Figure \ref{fig:fig1}(c), a polyelectrolyte chain at equilibrium relaxes at a much faster timescale (shorter Zimm time) compared to its translocation time in the range of chain lengths used and for typical voltages applied in these studies. Moreover, Figure \ref{fig:fig1}(d) shows that the chain relaxes at a faster timescale compared to the translocation time per Kuhn segment.
Based on these data, a polyelectrolyte chain initially at equilibrium can be assumed to be in quasi-equilibrium as it translocates across the nanopore (Figure \ref{fig:fig1}(a)). 

A one-dimensional free energy landscape can be constructed for the translocation process, using the equilibrium scaling theory for polymers \cite{Eisenriegler1982,Muthukumar2003} to calculate entropic contribution towards the free energy. All experimentally relevant parameters can be taken into consideration while constructing the free energy. Chemical details of the nanopore such as electrostatics and hydrophobicity are modeled using the effective pore-polymer interaction strength, while the salt concentration and pH in the two reservoirs can be accounted for in the entropic contribution. A model based on the Fokker-Planck (FP) formalism is used to describe the translocation process \cite{Sung1996,Muthukumar2003}. With the appropriate free energy landscape as an input, the FP equation can predict the mean translocation time and the translocation time distribution (histogram) with an effective friction coefficient as the only parameter. Excellent agreement with the model prediction  has been reported for translocation time histograms from simulations \cite{McCaffrey2013,Dunn2014,Katkar2014} and experiments \cite{Muthukumar2015}. 

In the Fokker-Planck formalism for single-file translocation, the entropic contribution towards the free energy landscape is derived by implying quasi-equilibrium assumption even for translocation of every segment. This is a stricter criterion for applicability of the quasi-equilibrium approximation than the above condition ($\tau_\text{Zimm} < \langle\tau\rangle$). As discussed below (Section \ref{sec:results}), even this criterion seems to be satisfied for the current experimental situations and simulations. Nevertheless, for high transmembrane voltages and for longer polymers the quasi-equilibrium assumption is expected to break down and an alternative approach is necessary. The fractional FP model takes into consideration the memory effects in the transport of the polyelectrolyte chain in terms of ``folds" (sections of the chain) that translocate by crossing entropic barriers \cite{Metzler2003,Dubbeldam2007,Dubbeldam2007a} giving rise to anomalous diffusion. A similar approach is developed by considering  time-dependent drift velocity and diffusion coefficient originating from velocity fluctuations to arrive at a modified FP equation \cite{Dubbeldam2013}. The tension propagation (TP) model \cite{Sakaue2007,Sakaue2010,Rowghanian2011,Rowghanian2012} assumes a two-state picture, whereby only a part of the chain near the nanopore entrance is stretched during early stages of translocation. The rest of the chain is stationary. The stretching of part of the chain results into formation of blobs of increasing size starting from the nanopore entrance (Figure \ref{fig:fig1}(b)). A balance between the driving force and the friction due to formation of blobs gives rise to the dynamic TP equation\cite{Sakaue2007}. The boundary separating the stretched and stationary parts of the chain propagates towards the trailing end according to this equation until it encompasses the entire chain. This gives rise to a time dependent friction coefficient resulting from the stretched part of the chain in the donor reservoir, in contrast to a constant effective monomer friction coefficient used in the FP model.

The predictions of the FP, fractional FP, and TP models on the dependence of $\langle\tau\rangle$ on $N$ and $V$ are quite varied and rich. A brief summary is as follows. While the predictions of the FP model have been validated by experimental results gathered so far, albeit for not very long polyelectrolyte chains, the predictions of the fractional FP and TP models remain untested. The FP model predicts that for a chain of length $N$ translocating under an applied voltage $V$, the mean translocation time scales as $\langle\tau\rangle \sim N/V$. The fractional FP model predicts a scaling \cite{Dubbeldam2007} of $\langle\tau\rangle \sim N^{1.5}/V$. TP model predicts different scaling regimes based on the force due to the applied voltage: $\langle\tau\rangle \sim N^{2\nu}/V$ for a weak force and $\langle\tau\rangle \sim N^{1+\nu}/V$ for extreme driving force. For intermediate driving forces, the scaling relation is $\langle\tau\rangle \sim N^{\frac{1+3\nu}{2}}/V^{\frac{3\nu-1}{2\nu}}$ if the velocity of the smallest blob is used as a velocity scale for TP \cite{Saito2011}; else $\langle\tau\rangle \sim N^{1+\nu}/V^{\frac{2\nu-1}{\nu}}$ if the velocity of the largest blob is chosen \cite{Saito2012err,Saito2012} or if a constant monomer flux is assumed in the stretched chain \cite{Rowghanian2011}.  Thus, the FP model predicts a uniform inverse dependence of the mean translocation time with voltage, while the TP model predicts three regimes- an inverse scaling with voltage at the two extremes along with an intermediate cross-over regime. Brownian dynamics simulations based on a time dependent friction derived from the TP model \cite{Ikonen2012} show that for short polymers and for narrow nanopores, a major part of the total friction comes from the friction with the nanopore \cite{Ikonen2012a,Ikonen2013}, making the contribution from the frictional blobs less relevant and leading to a scaling predicted by the FP model. These simulations use a high driving force, which corresponds to an equivalent experimental transmembrane voltage ranging from \cite{Ikonen2012a,Ikonen2013} $0.5$ V to \cite{Sean2015} $5$ V applied across a $5$ nm long nanopore, such as the $\alpha$-HL nanopore, whereas the applied voltage in the experiments is about 150 mV.

The primary focus of the present paper is to evaluate the role of non-equilibrium conformations of a polyelectrolyte chain in its translocation kinetics by simulating the average translation time and the relaxation time for a range of $V$. We report the effect of chain stretching on the kinetics of its translocation and compare the mean translocation time with a characteristic relaxation time for out-of-equilibrium chain conformations, obtained from coarse-grained Langevin dynamics simulations performed using LAMMPS package \cite{Plimpton1995}. We observe that even highly stretched initial chain conformations translocate at timescales comparable to their relaxation timescale. We also observe that the translocation time becomes less sensitive to initial chain conformations as the applied transmembrane voltage increases. Additionally, we observe a scaling of $\langle \tau \rangle \sim 1/V$ for the entire range of $V$ studied, which is in agreement with predictions of the FP model. We limit our study of translocation and relaxation of initially stretched chains to the parametric dependence on $V$, and exclude the dependence of the physically important but computationally expensive parameter $N$ from our simulations.  Instead, we use experimental results from the literature to set an upper bound on chain length below which the FP model is valid. Our simulation methodology is  discussed in Section \ref{sec:simulations}, and the corresponding results are discussed in Section \ref{sec:results}.

\section{Simulations}
\label{sec:simulations}
We adopt a two-step strategy in our coarse-grained simulations. In the first step, we perform simulations in presence of a fluid-flow to systematically generate a set of stretched chain conformations. In the next step, the same set of stretched chain conformations is subjected to two separate simulations in absence of the fluid-flow: translocation simulations and relaxation simulations.

For the first step, during which equilibrium polyelectrolyte chain conformations are stretched to systematically generate non-equilibrium conformations, we use hydrodynamics simulations performed using the multi-particle collision (MPC) dynamics technique. The coarse-grained polyelectrolyte chain is gradually stretched due to the underlying MPC fluid flowing through a cylindrical channel.  For the second step, we use Langevin dynamics simulations of a coarse-grained polyelectrolyte to study its translocation through a coarse-grained nanopore under the action of a constant electric field, acting along the nanopore axis, applied inside the nanopore. We also perform relaxation simulations in absence of the nanopore or the electric field, where no fluid flow is present. The details of each of these simulations, along with our coarse-grained model for the polyelectrolyte and the nanopore, are provided below. Unless noted otherwise, all quantities below are presented in reduced units using a length-scale of $3$ \AA\text{ } and an energy scale of $k_B T$, with $T=300$ K.

\subsection{Coarse-grained models}

\subsubsection*{Nanopore}
In translocation simulations, the system is made up of three components: an uncharged membrane, a cylindrical nanopore inside the membrane, and a charged polyelectrolyte chain. Details of each of these components are described below and are similar to those presented in Ref \cite{Katkar2014}, with difference in the choice of the length and energy scales as specified above.

The membrane is $16$ unit thick, and contains a cylindrical nanopore of radius $1.8$.  The membrane walls are modelled using spherical beads of diameter $d_1=1$ arranged into a two dimensional grid with grid-spacing of $d_1$. The nanopore wall is made up of smaller beads of diameter $d_2$, with the ratio $d_2/d_1 = 0.25$. These beads are placed along the circumference of a cylinder, with each bead approximately at a distance of $d_2/2$ from each other. This overlapping of nanopore wall beads is done to make the nanopore cross-section more circular. A part of the membrane walls surrounding both ends of the nanopore is made up of beads identical to the nanopore wall beads. Note that the nanopore is present only in the translocation simulations, and is absent in the relaxation simulations.

\subsubsection*{Polyelectrolyte}

$N$ beads of diameter $d_1$ are linearly connected by harmonic bonds to form a uniform polyelectrolyte chain. The equilibrium bond length is taken to be equal to $d_1$. A unit negative charge is placed on each polyelectrolyte bead.

\subsection*{Pair-wise interactions}
Excluded volume interactions are modelled using a truncated Lennard-Jones potential between two beads, given by the equation,
\begin{equation}
U_{\text{LJ}} = \left\{ \begin{array}{rl}
4 \epsilon_{\text{LJ}} \left[\left(\frac{\sigma}{r}\right)^{12} - \left(\frac{\sigma}{r}\right)^6\right] + \epsilon_{\text{LJ}} &\mbox{ for } r \leq 1.12 \sigma \\
0 & \mbox{ for } r > 1.12 \sigma.
\end{array}\right.
\end{equation}
Here, $r$ is the distance between two beads. $\epsilon_{\text{LJ}} = 1$ is the depth of the potential well. The potential is truncated at its minimum, corresponding to a distance of $1.12\sigma$. These interactions exist between all the beads, with $\sigma = 1$ for  interactions between polymer-polymer and polymer-membrane beads, and $\sigma = 0.625$ for nanopore-polymer beads. The positions of nanopore and membrane beads are fixed in the simulation and hence, pair-wise interactions are not computed for pairs of these beads.

Electrostatic interactions between a pair of beads are modelled using the truncated Debye-H\"{u}ckel potential, with an inverse of the Debye length $\kappa^{-1}=0.81$,
\begin{equation}
U_{\text{DH}} =  \frac{C q_i q_j}{\epsilon} \frac{\exp(-\kappa r)}{r}  \quad\quad\mbox{ for } r \leq r_{c2}
\end{equation}
with $q_i$ and $q_j$ corresponding to the charges on the two beads, separated by a distance $r$. Although the effective dielectric constant of an electrolyte solution confined within a nanopore is unknown, we have taken $\epsilon = 80$. $r_{c2}$ is the cutoff distance at which the electrostatic interactions are truncated. A value of $r_{c2} = 3$ is used in all the simulations.

Pair-wise electrostatic interactions are computed between pairs of polymer beads, using $q_1 = q_2 = -1e$, where $e$ is the elementary charge. As described in Ref.  \cite{Plimpton1995}, the energy conversion constant $C=1$ in reduced units.

The bonds between polymer beads are modelled using a harmonic potential. The equilibrium bond length between two connected polymer beads is $r_0 = 1$. The value of spring constant $K=15480$ is large enough to prevent unrealistic bond extensions..
\begin{equation}
U_{b} = K(r-r_0)^2
\end{equation}

\subsection{Langevin dynamics}

In Langevin dynamics simulations, the above potentials are used to compute forces on each of the polyelectrolyte beads. The equation of motion for each bead is given by
\begin{equation}
m \frac{d^2r}{dt^2} = - \zeta \frac{dr}{dt} - \nabla (U_{\text{LJ}} + U_{\text{DH}} + U_{b}) + F_{r} + F_{\text{ext}}
\end{equation}
where $m$ is mass of the bead, $\zeta$ is the friction coefficient, and $F_{r} \sim \sqrt{k_B T \zeta/dt}$ is the random force due to solvent at the given temperature $T$, as documented extensively in the literature \cite{Plimpton1995}.  We choose the values $m=1$ and $\zeta=1$ in all our simulations. 

The force due to applied electric field $E$ acting on each bead is given by $F_{\text{ext}} = q_1 E$. The electric field is assumed to be constant, acts along the axis of the nanopore, and is present only inside the nanopore.

\subsection{Multi-particle collision dynamics}

The mesoscale multi-particle collision (MPC) dynamics technique is used to systematically generate stretched polyelectrolyte chain conformations by subjecting them to flow in a cylindrical channel. The details of this technique can be found in Refs. \cite{Malevanets1999,Malevanets2000,Ihle2001}. Briefly, the solvent is modeled using particles with their dynamics governed by two steps, streaming and collision, that provide hydrodynamics correlations between the solute molecules in addition to flow fields. Three parameters of the model decide the solvent properties, the particle density $\rho_\text{MPC}$, the timestep for a streaming step $h_\text{MPC}$, and the rotation angle for the collision step $\theta_\text{MPC}$. The simulation box is divided into unit cubic cells, and $\rho_\text{MPC}$ is the number of solvent particles in a unit cell. We choose $\rho_\text{MPC}=10$, $h_\text{MPC}=0.1$ and $\theta_\text{MPC} = 130^\circ$, mimicking a fluid with a viscosity of $8.7$.  During the streaming step of duration $h_\text{MPC}$, each solvent particle moves with a constant acceleration of $0.0005$ in the direction of flow along the axis of the channel and a constant velocity in the two other directions, without interacting with other particles. Bounce-back rule is used to apply no-slip condition at the channel surface. For a polymer length of $N=120$, the channel radius is set at $27.5$, and periodic boundary condition is used in the direction of flow.  The solute particles, i.e. the polyelectrolyte beads, undergo regular molecular dynamics during the streaming step, and interact with the solvent by participating in the collision step. During the collision step, the relative momenta of the solvent and solute particles belonging to a cell are instantaneously rotated along a random direction with the angle $\theta_\text{MPC}$. The mass of the solute particles is assumed to be 10\% of the mass of a polyelectrolyte bead.

In summary, the polyelectrolyte chain is represented using an array of $120$ negatively charged beads of size $1$ connected linearly by harmonic springs, with Debye-H\"{u}ckel electrostatics corresponding to a Debye length of $1.23$. Simulation parameters such as the friction coefficient affect the experimental regime that the simulations mimic, and should be carefully chosen to avoid simulation artifacts and to simulate the appropriate balance between drift and diffusion. In our Langevin dynamics simulations, we choose a friction coefficient of $1$ and vary the  driving electric force in the range of $0.24$ to $1.67$ per bead, which has been shown to be an appropriate choice for mimicking the translocation of a ssDNA with $120$ beads \cite{deHaan2015}.
\begin{figure}[t]
	\centering
		\includegraphics*[width=\myfigwidth]{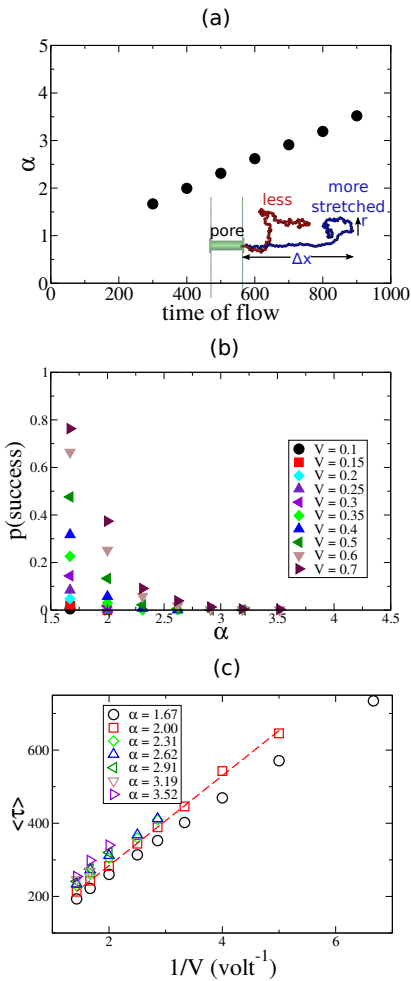}
		\caption{(a) Effect of time of flow on the average aspect ratio $\alpha=\Delta x/2r$ of the chain. Inset shows a snapshot of the coarse-grained translocation setup with two stretched chain conformations overlaid, and the definitions of $\Delta x$ and $r$. (b) Probability of successful translocation and (c) mean translocation time $\langle\tau\rangle$ as a function of $\alpha$ and $V$ (volts) applied across the 4.8 nm long nanopore. The dotted red line is a sample linear fit emphasizing the scaling of $\langle\tau\rangle \sim 1/V$. }
		\label{fig:fig2}
\end{figure}
\subsection{Generating stretched conformations}
To deliberately generate stretched polymer chain conformations, we first generate an equilibrium conformation by performing Langevin dynamics simulation of a chain near a wall for 40000 time units, with one end of the chain fixed near the wall. In the next step, the final conformation from the previous step is subjected to fluid flow inside a cylindrical channel using hydrodynamics simulations \cite{Malevanets1999} implemented using the MPC dynamics technique, with one end of the chain fixed.  Chain conformations with increasing levels of stretching are systematically generated based on the time that the chain is subjected to flow. An average aspect ratio $\alpha=\Delta x/(2 r)$ is used to characterize each set of chain conformations, which monotonically increases with the time of flow (Figure \ref{fig:fig2}(a)). $\Delta x$ is the maximum span of the chain in the $x$ direction and $r$ is its maximum radius ($r=\sqrt{y_{max}^2+z_{max}^2}$) from its center of mass in the $y$-$z$ plane (inset Figure \ref{fig:fig2}(a)).

\subsection{Translocation simulations}
A $16$ units long uncharged cylindrical nanopore with a radius of $1.8$ is used for translocation simulations (dimensions based on an $\alpha$-HL nanopore, inset Figure \ref{fig:fig2}(a)). A constant electric field equal to $V/16$ is applied inside the nanopore. The polyelectrolyte chain undergoes Langevin dynamics, in absence of any fluid-flow, and translocates across the nanopore under the action of the applied electric field. Mean translocation time $\langle \tau \rangle$ is defined as the average time taken by the polyelectrolyte chain, initially at the nanopore entrance, to exit through the nanopore into the receiver reservoir (successful translocation), and is obtained from 2500 statistically independent simulation runs for each combination of $\alpha$ and $V$. Probability of successful translocations is calculated as the fraction of the runs that are successful.

\subsection{Relaxation simulations}
The relaxation simulations are performed in absence of the nanopore and the electric field. Each polyelectrolyte chain with a conformation corresponding to a given level of stretching $\alpha$ undergoes Langevin dynamics in absence of any fluid-flow for sufficiently long time (16000 time units) until the initially stretched chain relaxes to an equilibrium state. An average moment of inertia tensor is accumulated from the 2500 independent simulation runs by re-centering each conformation using its center of mass, and average principal moments of inertia $\lambda_i$ (with i=1,2,3) computed from this tensor are used as a metric of relaxation. As a reminder, a conformation corresponding to a shape with one symmetric axis is characterized by a smaller moment of inertia along the axis ($\lambda_1$) and two large and identical moments of inertia ($\lambda_2 = \lambda_3$). A spherically symmetric object, such as the conformation of a polyelectrolyte when it is completely relaxed, has all three principal moments of inertia identical ($\lambda_1 = \lambda_2 = \lambda_3$).

\begin{figure}
	\centering
		\includegraphics*[width=\myfigwidth]{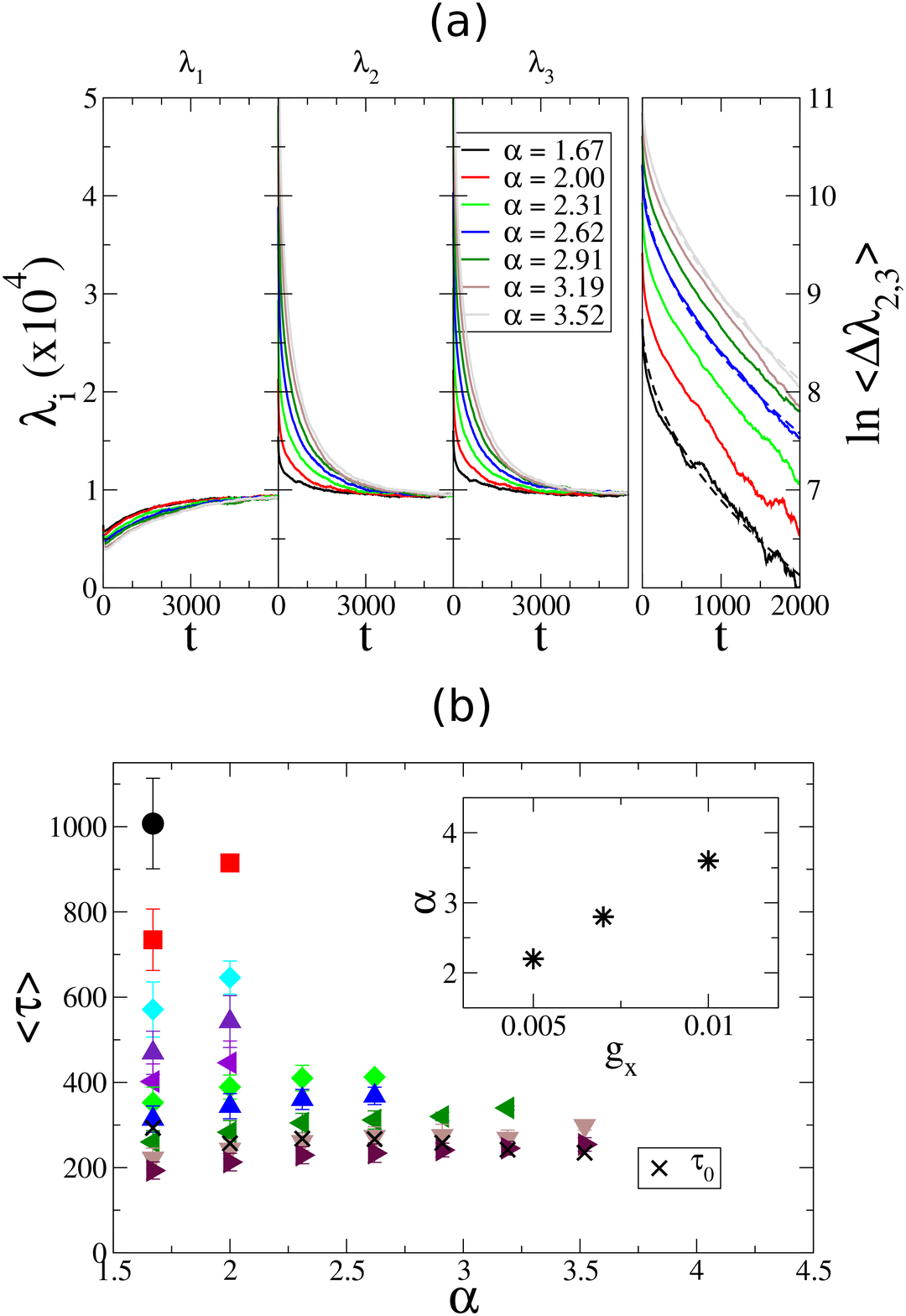}
		\caption{(a) The average principal moments of inertia $\lambda_i$ of a stretched polyelectrolyte chain undergoing relaxation (left three panels). A log-linear plot showing the stretched exponential initial decay of $\lambda_2$ and $\lambda_3$, followed by a systematic exponential decay is also shown (right panel). The dotted lines are fits showing the stretched exponential behavior at shorter times.  (b) Comparison of relaxation timescale $\tau_0$ and translocation time $\langle\tau\rangle$ for stretched chains. Filled symbols are similar to those in Figure \ref{fig:fig2}(b). Inset shows the strength of fluid flow required to achieve equivalent chain stretching $\alpha$.}
				\label{fig:fig3}
\end{figure}

\section{Results and discussion}
\label{sec:results}
As stretching of the initial conformation of the chain increases, it is more likely to retract back into the donor compartment than to translocate across the nanopore. This is reflected in the probability of successful translocations that decreases with increase in $\alpha$ for all applied voltages (Figure \ref{fig:fig2}(b)). The pull from the initially stretched entropic spring becomes relatively stronger in comparison with the driving electric field, making the probability of successful translocations nearly zero beyond a certain level of stretching for a given voltage (\textit{e.g.}, beyond $\alpha=2.31$ for $V=0.35$ V). The mean translocation time $\langle\tau\rangle$ increases with an increase in initial stretching, since the pull from the entropic spring has to be overcome by the applied electric field, causing a delay in $\langle\tau\rangle$ (Figure \ref{fig:fig2}(c)). Note that as the probability of translocation approaches zero, the uncertainty in estimating $\langle\tau\rangle$ from simulations increases. The increase in $\langle \tau \rangle$ with $\alpha$ becomes less prominent as the applied voltage increases. For example, at $V=0.2$ V, $\langle \tau \rangle$ varies in the range of $571-646$ over a narrow range of $\alpha$ between $1.67-2$, while at $V=0.7$ V, $\langle \tau \rangle$ varies in the range of $193-254$ over a much wider range of $\alpha$ between $1.67-3.52$. A more significant variation in $\langle\tau\rangle$ is observed with respect to the applied voltage. Specifically, a scaling of $\langle\tau\rangle \sim 1/V$ is observed for each level of stretching, similar to the scaling reported from simulations of unstretched polyelectrolytes in the literature \cite{Ikonen2012a,Adhikari2013}. Remarkably, the observed scaling with voltage is in excellent agreement with the prediction of the FP model, albeit the latter is developed within the quasi-equilibrium assumption. The TP model predicts the same scaling with voltage, but only for very weak ($V=0.0015$ V) and very strong ($V=0.457$ V) driving forces \cite{Saito2011}, with a transition to a sub-linear ($\langle\tau\rangle \sim 1/V^{2/3}$ in Ref. \cite{Saito2011} or $\langle\tau\rangle \sim 1/V^{1/3}$ in Refs. \cite{Saito2012err,Saito2012,Rowghanian2011}) scaling in the intermediate voltage regime. We do not observe any of these intermediate scaling regimes predicted by the TP model or any departure from inverse scaling with voltage even for highly stretched initial chain conformations used in our simulations.

As a stretched chain undergoes relaxation, the initially large difference between its principal moments $\lambda_1$ (along $x$-direction) and $\lambda_2\sim\lambda_3$ (in the $y$-$z$ plane) systematically decreases until all three become equal, corresponding to a spherically symmetric chain conformation (Figure \ref{fig:fig3}(a)). The mean of $\lambda_2$ and $\lambda_3$ relative to their value when the chain is completely relaxed is defined as $\langle\Delta\lambda_{2,3}\rangle = \sum_{i=2,3} (\lambda_i(t)-\lambda_i^\infty)/2$, where $\lambda_i^\infty$ is the average value of $\lambda_i$ for $t>8000$.  We fit the data using a stretched exponential decay of the form $\langle\Delta\lambda_{2,3}\rangle \sim e^{-\left[t/\tau_0\right]^{0.5}}$ to extract $\tau_0$, the characteristic decay time \cite{Cherayil1992} (Figure \ref{fig:fig3}(a)). A comparison of the mean translocation time from the translocation simulations with the relaxation time $\tau_0$ calculated from the relaxation simulations is shown in Figure \ref{fig:fig3}(b). For all levels of stretching, the relaxation timescale $\tau_0$ is nearly constant. Slight variations in $\tau_0$ are due to the specific range used in fitting the stretched exponential in Figure \ref{fig:fig3}(a). More important is the observation that the relaxation timescale is of the same order of magnitude as the mean translocation time in the ranges of initial stretching and voltage studied. As discussed earlier, the voltage applied across a $5$ nm long $\alpha$-HL nanopore in typical experiments is in the range of $0.04$-$0.26$ V. $\tau_0$ and $\langle\tau\rangle$ are quantitatively comparable for this experimentally relevant voltage range even for highly stretched chains studied here. To emphasize that the chains used in this study are, in fact, highly stretched, we measure the strength of fluid flow required to stretch the chains by performing hydrodynamics simulations \cite{Malevanets1999} of a polyelectrolyte chain in a simpler quasi-two dimensional setup consisting of fluid flow between parallel plates using the MPC dynamics technique. The electro-osmotic flow rate required in order to obtain such highly stretched chains (Inset Figure \ref{fig:fig3}(b)) is $\sim O(100)$ times stronger than its typical value of $35$ cm/s predicted in a $15$ nm wide nanopore \cite{Wong2007}. Moreover, $\tau_0$ reported here is the time taken for an initially stretched polyelectrolyte chain (\textit{e.g.}, due to an electro-osmotic flow field) to relax to an equilibrium state in absence of any such flow. Although a polyelectrolyte chain translocating through a nanopore could potentially undergo stretching due to an external field, such a field would not abruptly switch off as translocation progresses. Hence, $\tau_0$ should, at the best, be interpreted as an upper bound of the true relaxation timescale.

For weak driving forces (lower transmembrane voltages), a sufficient criterion for the quasi-equilibrium assumption to be valid during translocation is that the average time required for a unit translocation step is higher than the Zimm relaxation time. In that case, one can safely use pre-averaging over polymer chain conformations to calculate the entropic contribution to the total free energy using equilibrium theory for polymers. Considering the displacement of the polymer by Kuhn length $\ell$ as the unit step in translocation, one can re-plot the mean translocation time data obtained from experiments (Figure \ref{fig:fig1}(c)) in terms of the average time taken for a unit translocation step, and compare it with the Zimm relaxation time. Figure \ref{fig:fig1}(d) shows the same experimental data as Figure \ref{fig:fig1}(c), re-plotted using the above definition of a unit translocation step with $\ell=2$ nm. It is remarkable that all data corresponding to NaPSS, dextran sulfate sodium and ss-DNA fall well above the Zimm relaxation time, in-spite of different experimental conditions such as the applied voltage, salt concentration, temperature, etc. Thus, the sufficient criterion for the quasi-equilibrium assumption appears to be satisfied in all these experiments.

In contrast, the sufficient criterion for the quasi-equilibrium assumption is not satisfied in our simulations of initially stretched chains. The initially out-of-equilibrium chain remains out of equilibrium as it undergoes several translocation steps. However, the observed scaling of mean translocation time with the applied voltage in our simulations suggests that the non-equilibrium effects are not strong enough to deviate from predictions of the FP theory. The scaling of translocation kinetics with chain length, predicted to be different across different theories, is excluded from the present work owing to the computational costs involved, and is left to be studied in the future.

A prominent exception in the experimental results is the data set from Ref. \cite{Wanunu2008}. The data in Figure \ref{fig:fig1}(d) corresponding to these experiments differ from the data in all other experiments in terms of the nanopore ($10$ nm long, $3.5-4$ nm wide SiN nanopore) and the polyelectrolyte (ds-DNA, with $\ell=37.5$ nm), along with other experimental conditions such as the applied voltage ($200-500$ mV for the data-points positioned vertically at $N_\text{K} \sim 2.67 \Rightarrow N = 400$ bp, $300$ mV for others). More importantly, the data analysis used to calculate the mean translocation time from translocation events is different in these experiments. The total population of translocation events is divided into fast and slow classes. In Figure \ref{fig:fig1}(d), we include the data for the slow class as well. A Kuhn length of $37.5$ nm is used for dsDNA (corresponding to $1$ M salt concentration used in these experiments). Most of this data corresponds to the limit of $N_\text{K} \sim O(1)$, whereas Equation \ref{eqn:Zimm} is accurate only in the limit of large $N_\text{K}$. Keeping in mind the inapplicability of the Zimm theory for such short polymers, we avoid showing the corresponding Zimm time in the inset of Figure \ref{fig:fig1}(d). Nonetheless, a na\"{i}ve extrapolation of Zimm relaxation time to the limit of $N_\text{K} \sim O(1)$ should reveal that the average time for a unit step in translocation of a polymer longer than a few Kuhn segments is faster than the extrapolated $\tau_\text{Zimm}$. Thus, the sufficient condition for applicability of the quasi-equilibrium assumption is bound to be violated for dsDNA longer than a few Kuhn segments, under these experimental conditions.

It is important to realize that as the driving force increases, the entropic contribution to the free energy becomes increasingly less significant, especially for polymers that are significantly longer than the nanopore. Using equilibrium estimates, the entropic penalty of placing a Kuhn segment into the nanopore is of the order of $k_B T (\log z + 0.6 \log N_\text{K}/(N_\text{K}-1))\approx 2 k_B T$ ($z$ being the coordination number \cite{Muthukumar2011}), which is equivalent to the electrochemical energy $=eV/k_BT$ gained by a unit charge successfully translocating across a nanopore with $50$ mV applied transmembrane voltage at room temperature. To put this in perspective, a transmembrane voltage as small as $20$ mV already contributes to an electrochemical energy gain of $2k_B T$ for every Kuhn segment translocating across the nanopore, assuming a modest linear charge density of $1.33$ charges per nm of the polyelectrolyte (equivalent to a degree of ionization $\alpha_p=1/3$ for NaPSS). Although these estimates could slightly vary depending on the polyelectrolyte density and the estimate used for the entropic contribution, for polyelectrolytes that are significantly longer than the nanopore and are close to equilibrium as translocation begins, the electrochemical contribution to the free energy begins to dominate over the entropic contribution even at small transmembrane voltages. Pore-polymer interactions, electric field inside the nanopore (in addition to the electrochemical energy) further contribute to the free energy during certain stages of translocation, making the entropic contribution even less significant.

For short polymers that are only moderately longer than the length of the nanopore, the time required for filling the nanopore with the polymer chain can significantly contribute to the total translocation time. The entropic contribution to the free energy should then be compared to the gain in free energy due to electric field inside the nanopore, since the electrochemical energy does not contribute during this stage. For the first Kuhn segment to enter the nanopore, the entropic penalty of $\approx 2 k_B T$ has to be overcome by the energy due to electric field. Again, assuming a linear charge density of $1.33$ charges per nm of the polyelectrolyte, the required voltage applied across a $5$ nm long nanopore to overcome the entropic penalty is $\approx 100$ mV. Once the first Kuhn segment is inside the nanopore, the second Kuhn segment needs only $\approx 50$ mV voltage across the nanopore to match the net entropic loss. For higher polyelectrolyte charge density, the required voltage is even lower. Thus, as the applied voltage increases, the entropic contribution becomes less important.

Given a particular set of experimental conditions, the time taken for the polymer segment to translocate by a Kuhn length is longer than the Zimm time only upto a certain polymer length. Hence, the sufficient condition for quasi-equilibrium assumption is not satisfied for an infinitely long polymer. The polymer length at which the cross-over happens depends on the applied voltage. The translocation time scales with the applied voltage as $\langle\tau\rangle \sim 1/V^\beta$, with $\beta>0$ across all theories of driven polymer translocation.  As the applied voltage increases, the quasi-equilibrium assumption is violated at relatively shorter polymer lengths. 

In-spite of the relatively high polyelectrolyte charge density (a unit charge per monomer), high initial stretching of the chain, and a modest chain length used in our simulations, we already observe the diminishing contribution of polymer entropy (decreasing sensitivity of the mean translocation time towards the level of initial chain stretching) as the transmembrane voltage increases (Figures \ref{fig:fig2}(c) and \ref{fig:fig3}(b)). These data indicate that for high transmembrane voltages, the initial chain conformation and hence the chain entropy has negligible effect on translocation kinetics.

In summary, the translocation and relaxation timescales for a polyelectrolyte chain translocating through a nanopore are found to be comparable even for the out-of-equilibrium initial chain configurations studied, especially for experimentally relevant voltages. Although the Fokker-Planck model is able to describe the rich phenomenology in polymer translocation and is widely used to gain insights into translocation experiments \cite{Meller2001,Hoogerheide2013,Muthukumar2015}, it would be inadequate for very large molecules and very high voltages. Our work establishes the range of applicability of the Fokker-Planck model and the onset of non-equilibrium effects for flexible polyelectrolytes at various experimentally relevant voltage ranges. Extensions of the present study of significantly stretched initial conformations to different chain lengths over several orders of magnitude, and an assessment of non-equilibrium chain conformations while the chain is undergoing translocation under experimentally relevant conditions are of further interest.

 \begin{acknowledgments}
 	Acknowledgement is made to the National Institutes of Health (Grant No. R01HG002776-11), National Science Foundation (Grant No. 1504265) and Air Force Office of Scientific Research (Grant No. FA9550-14-1-0164).
 \end{acknowledgments}


\begin{thebibliography}{64}%
\makeatletter
\providecommand \@ifxundefined [1]{%
 \@ifx{#1\undefined}
}%
\providecommand \@ifnum [1]{%
 \ifnum #1\expandafter \@firstoftwo
 \else \expandafter \@secondoftwo
 \fi
}%
\providecommand \@ifx [1]{%
 \ifx #1\expandafter \@firstoftwo
 \else \expandafter \@secondoftwo
 \fi
}%
\providecommand \natexlab [1]{#1}%
\providecommand \enquote  [1]{``#1''}%
\providecommand \bibnamefont  [1]{#1}%
\providecommand \bibfnamefont [1]{#1}%
\providecommand \citenamefont [1]{#1}%
\providecommand \href@noop [0]{\@secondoftwo}%
\providecommand \href [0]{\begingroup \@sanitize@url \@href}%
\providecommand \@href[1]{\@@startlink{#1}\@@href}%
\providecommand \@@href[1]{\endgroup#1\@@endlink}%
\providecommand \@sanitize@url [0]{\catcode `\\12\catcode `\$12\catcode
  `\&12\catcode `\#12\catcode `\^12\catcode `\_12\catcode `\%12\relax}%
\providecommand \@@startlink[1]{}%
\providecommand \@@endlink[0]{}%
\providecommand \url  [0]{\begingroup\@sanitize@url \@url }%
\providecommand \@url [1]{\endgroup\@href {#1}{\urlprefix }}%
\providecommand \urlprefix  [0]{URL }%
\providecommand \Eprint [0]{\href }%
\providecommand \doibase [0]{http://dx.doi.org/}%
\providecommand \selectlanguage [0]{\@gobble}%
\providecommand \bibinfo  [0]{\@secondoftwo}%
\providecommand \bibfield  [0]{\@secondoftwo}%
\providecommand \translation [1]{[#1]}%
\providecommand \BibitemOpen [0]{}%
\providecommand \bibitemStop [0]{}%
\providecommand \bibitemNoStop [0]{.\EOS\space}%
\providecommand \EOS [0]{\spacefactor3000\relax}%
\providecommand \BibitemShut  [1]{\csname bibitem#1\endcsname}%
\let\auto@bib@innerbib\@empty
\bibitem [{\citenamefont {Kasianowicz}\ \emph {et~al.}(1996)\citenamefont
  {Kasianowicz}, \citenamefont {Brandin}, \citenamefont {Branton},\ and\
  \citenamefont {Deamer}}]{Kasianowicz1996}%
  \BibitemOpen
  \bibfield  {author} {\bibinfo {author} {\bibfnamefont {J.~J.}\ \bibnamefont
  {Kasianowicz}}, \bibinfo {author} {\bibfnamefont {E.}~\bibnamefont
  {Brandin}}, \bibinfo {author} {\bibfnamefont {D.}~\bibnamefont {Branton}}, \
  and\ \bibinfo {author} {\bibfnamefont {D.~W.}\ \bibnamefont {Deamer}},\
  }\href {http://www.pnas.org/content/93/24/13770} {\bibfield  {journal}
  {\bibinfo  {journal} {P. Natl. Acad. Sci. U.S.A.}\ }\textbf {\bibinfo
  {volume} {93}},\ \bibinfo {pages} {13770} (\bibinfo {year}
  {1996})}\BibitemShut {NoStop}%
\bibitem [{\citenamefont {Muthukumar}(2011)}]{Muthukumar2011}%
  \BibitemOpen
  \bibfield  {author} {\bibinfo {author} {\bibfnamefont {M.}~\bibnamefont
  {Muthukumar}},\ }\href@noop {} {\emph {\bibinfo {title} {Polymer
  Translocation}}}\ (\bibinfo  {publisher} {Boca Raton: Taylor \& Francis},\
  \bibinfo {year} {2011})\BibitemShut {NoStop}%
\bibitem [{\citenamefont {Wanunu}\ \emph {et~al.}(2010)\citenamefont {Wanunu},
  \citenamefont {Morrison}, \citenamefont {Rabin}, \citenamefont {Grosberg},\
  and\ \citenamefont {Meller}}]{Wanunu2010}%
  \BibitemOpen
  \bibfield  {author} {\bibinfo {author} {\bibfnamefont {M.}~\bibnamefont
  {Wanunu}}, \bibinfo {author} {\bibfnamefont {W.}~\bibnamefont {Morrison}},
  \bibinfo {author} {\bibfnamefont {Y.}~\bibnamefont {Rabin}}, \bibinfo
  {author} {\bibfnamefont {A.~Y.}\ \bibnamefont {Grosberg}}, \ and\ \bibinfo
  {author} {\bibfnamefont {A.}~\bibnamefont {Meller}},\ }\href {\doibase
  10.1038/nnano.2009.379} {\bibfield  {journal} {\bibinfo  {journal} {Nat.
  Nanotechnol.}\ }\textbf {\bibinfo {volume} {5}},\ \bibinfo {pages} {160}
  (\bibinfo {year} {2010})}\BibitemShut {NoStop}%
\bibitem [{\citenamefont {Wong}\ and\ \citenamefont
  {Muthukumar}(2010)}]{Wong2010}%
  \BibitemOpen
  \bibfield  {author} {\bibinfo {author} {\bibfnamefont {C.~T.~A.}\
  \bibnamefont {Wong}}\ and\ \bibinfo {author} {\bibfnamefont {M.}~\bibnamefont
  {Muthukumar}},\ }\href {\doibase 10.1063/1.3464333} {\bibfield  {journal}
  {\bibinfo  {journal} {J. Chem. Phys.}\ }\textbf {\bibinfo {volume} {133}},\
  \bibinfo {pages} {045101} (\bibinfo {year} {2010})}\BibitemShut {NoStop}%
\bibitem [{\citenamefont {Rowghanian}\ and\ \citenamefont
  {Grosberg}(2013{\natexlab{a}})}]{Rowghanian2013}%
  \BibitemOpen
  \bibfield  {author} {\bibinfo {author} {\bibfnamefont {P.}~\bibnamefont
  {Rowghanian}}\ and\ \bibinfo {author} {\bibfnamefont {A.~Y.}\ \bibnamefont
  {Grosberg}},\ }\href {\doibase 10.1103/physreve.87.042722} {\bibfield
  {journal} {\bibinfo  {journal} {Phys. Rev. E}\ }\textbf {\bibinfo {volume}
  {87}},\ \bibinfo {pages} {042722} (\bibinfo {year}
  {2013}{\natexlab{a}})}\BibitemShut {NoStop}%
\bibitem [{\citenamefont {Rowghanian}\ and\ \citenamefont
  {Grosberg}(2013{\natexlab{b}})}]{Rowghanian2013a}%
  \BibitemOpen
  \bibfield  {author} {\bibinfo {author} {\bibfnamefont {P.}~\bibnamefont
  {Rowghanian}}\ and\ \bibinfo {author} {\bibfnamefont {A.~Y.}\ \bibnamefont
  {Grosberg}},\ }\href {\doibase 10.1103/physreve.87.042723} {\bibfield
  {journal} {\bibinfo  {journal} {Phys. Rev. E}\ }\textbf {\bibinfo {volume}
  {87}},\ \bibinfo {pages} {042723} (\bibinfo {year}
  {2013}{\natexlab{b}})}\BibitemShut {NoStop}%
\bibitem [{\citenamefont {Jeon}\ and\ \citenamefont
  {Muthukumar}(2014{\natexlab{a}})}]{Jeon2014}%
  \BibitemOpen
  \bibfield  {author} {\bibinfo {author} {\bibfnamefont {B.-j.}\ \bibnamefont
  {Jeon}}\ and\ \bibinfo {author} {\bibfnamefont {M.}~\bibnamefont
  {Muthukumar}},\ }\href {\doibase 10.1063/1.4855075} {\bibfield  {journal}
  {\bibinfo  {journal} {J. Chem. Phys.}\ }\textbf {\bibinfo {volume} {140}},\
  \bibinfo {pages} {015101} (\bibinfo {year} {2014}{\natexlab{a}})}\BibitemShut
  {NoStop}%
\bibitem [{\citenamefont {Cohen}, \citenamefont {Chaudhuri},\ and\
  \citenamefont {Golestanian}(2012)}]{Cohen2012}%
  \BibitemOpen
  \bibfield  {author} {\bibinfo {author} {\bibfnamefont {J.~A.}\ \bibnamefont
  {Cohen}}, \bibinfo {author} {\bibfnamefont {A.}~\bibnamefont {Chaudhuri}}, \
  and\ \bibinfo {author} {\bibfnamefont {R.}~\bibnamefont {Golestanian}},\
  }\href {\doibase 10.1103/physrevx.2.021002} {\bibfield  {journal} {\bibinfo
  {journal} {Phys. Rev. X}\ }\textbf {\bibinfo {volume} {2}},\ \bibinfo {pages}
  {238102} (\bibinfo {year} {2012})}\BibitemShut {NoStop}%
\bibitem [{\citenamefont {Katkar}\ and\ \citenamefont
  {Muthukumar}(2014)}]{Katkar2014}%
  \BibitemOpen
  \bibfield  {author} {\bibinfo {author} {\bibfnamefont {H.~H.}\ \bibnamefont
  {Katkar}}\ and\ \bibinfo {author} {\bibfnamefont {M.}~\bibnamefont
  {Muthukumar}},\ }\href {\doibase 10.1063/1.4869862} {\bibfield  {journal}
  {\bibinfo  {journal} {J. Chem. Phys.}\ }\textbf {\bibinfo {volume} {140}},\
  \bibinfo {pages} {135102} (\bibinfo {year} {2014})}\BibitemShut {NoStop}%
\bibitem [{\citenamefont {Banerjee}\ \emph {et~al.}(2014)\citenamefont
  {Banerjee}, \citenamefont {Wilson}, \citenamefont {Shim}, \citenamefont
  {Shankla}, \citenamefont {Corbin}, \citenamefont {Aksimentiev},\ and\
  \citenamefont {Bashir}}]{Banerjee2014}%
  \BibitemOpen
  \bibfield  {author} {\bibinfo {author} {\bibfnamefont {S.}~\bibnamefont
  {Banerjee}}, \bibinfo {author} {\bibfnamefont {J.}~\bibnamefont {Wilson}},
  \bibinfo {author} {\bibfnamefont {J.}~\bibnamefont {Shim}}, \bibinfo {author}
  {\bibfnamefont {M.}~\bibnamefont {Shankla}}, \bibinfo {author} {\bibfnamefont
  {E.~A.}\ \bibnamefont {Corbin}}, \bibinfo {author} {\bibfnamefont
  {A.}~\bibnamefont {Aksimentiev}}, \ and\ \bibinfo {author} {\bibfnamefont
  {R.}~\bibnamefont {Bashir}},\ }\href {\doibase 10.1002/adfm.201403719}
  {\bibfield  {journal} {\bibinfo  {journal} {Adv. Funct. Mater.}\ }\textbf
  {\bibinfo {volume} {25}},\ \bibinfo {pages} {936} (\bibinfo {year}
  {2014})}\BibitemShut {NoStop}%
\bibitem [{\citenamefont {Izmitli}\ \emph {et~al.}(2008)\citenamefont
  {Izmitli}, \citenamefont {Schwartz}, \citenamefont {Graham},\ and\
  \citenamefont {de~Pablo}}]{Izmitli2008}%
  \BibitemOpen
  \bibfield  {author} {\bibinfo {author} {\bibfnamefont {A.}~\bibnamefont
  {Izmitli}}, \bibinfo {author} {\bibfnamefont {D.~C.}\ \bibnamefont
  {Schwartz}}, \bibinfo {author} {\bibfnamefont {M.~D.}\ \bibnamefont
  {Graham}}, \ and\ \bibinfo {author} {\bibfnamefont {J.~J.}\ \bibnamefont
  {de~Pablo}},\ }\href {\doibase 10.1063/1.2831777} {\bibfield  {journal}
  {\bibinfo  {journal} {J. Chem. Phys.}\ }\textbf {\bibinfo {volume} {128}},\
  \bibinfo {pages} {085102} (\bibinfo {year} {2008})}\BibitemShut {NoStop}%
\bibitem [{\citenamefont {Saito}\ and\ \citenamefont
  {Sakaue}(2012{\natexlab{a}})}]{Saito2012}%
  \BibitemOpen
  \bibfield  {author} {\bibinfo {author} {\bibfnamefont {T.}~\bibnamefont
  {Saito}}\ and\ \bibinfo {author} {\bibfnamefont {T.}~\bibnamefont {Sakaue}},\
  }\href {\doibase 10.1103/PhysRevE.85.061803} {\bibfield  {journal} {\bibinfo
  {journal} {Phys. Rev. E}\ }\textbf {\bibinfo {volume} {85}},\ \bibinfo
  {pages} {061803} (\bibinfo {year} {2012}{\natexlab{a}})}\BibitemShut
  {NoStop}%
\bibitem [{\citenamefont {Sarabadani}, \citenamefont {Ikonen},\ and\
  \citenamefont {Ala-Nissila}(2014)}]{Sarabadani2014}%
  \BibitemOpen
  \bibfield  {author} {\bibinfo {author} {\bibfnamefont {J.}~\bibnamefont
  {Sarabadani}}, \bibinfo {author} {\bibfnamefont {T.}~\bibnamefont {Ikonen}},
  \ and\ \bibinfo {author} {\bibfnamefont {T.}~\bibnamefont {Ala-Nissila}},\
  }\href {\doibase 10.1063/1.4903176} {\bibfield  {journal} {\bibinfo
  {journal} {J. Chem. Phys.}\ }\textbf {\bibinfo {volume} {141}},\ \bibinfo
  {pages} {214907} (\bibinfo {year} {2014})}\BibitemShut {NoStop}%
\bibitem [{\citenamefont {Farahpour}\ \emph {et~al.}(2013)\citenamefont
  {Farahpour}, \citenamefont {Maleknejad}, \citenamefont {Varnik},\ and\
  \citenamefont {Ejtehadi}}]{Farahpour2013}%
  \BibitemOpen
  \bibfield  {author} {\bibinfo {author} {\bibfnamefont {F.}~\bibnamefont
  {Farahpour}}, \bibinfo {author} {\bibfnamefont {A.}~\bibnamefont
  {Maleknejad}}, \bibinfo {author} {\bibfnamefont {F.}~\bibnamefont {Varnik}},
  \ and\ \bibinfo {author} {\bibfnamefont {M.~R.}\ \bibnamefont {Ejtehadi}},\
  }\href {\doibase 10.1039/c2sm27416g} {\bibfield  {journal} {\bibinfo
  {journal} {Soft Matter}\ }\textbf {\bibinfo {volume} {9}},\ \bibinfo {pages}
  {2750} (\bibinfo {year} {2013})}\BibitemShut {NoStop}%
\bibitem [{\citenamefont {Vollmer}\ and\ \citenamefont
  {de~Haan}(2016)}]{Vollmer2016}%
  \BibitemOpen
  \bibfield  {author} {\bibinfo {author} {\bibfnamefont {S.~C.}\ \bibnamefont
  {Vollmer}}\ and\ \bibinfo {author} {\bibfnamefont {H.~W.}\ \bibnamefont
  {de~Haan}},\ }\href {\doibase 10.1063/1.4964630} {\bibfield  {journal}
  {\bibinfo  {journal} {J. Chem. Phys.}\ }\textbf {\bibinfo {volume} {145}},\
  \bibinfo {pages} {154902} (\bibinfo {year} {2016})}\BibitemShut {NoStop}%
\bibitem [{\citenamefont {Jeon}\ and\ \citenamefont
  {Muthukumar}(2014{\natexlab{b}})}]{Jeon2014a}%
  \BibitemOpen
  \bibfield  {author} {\bibinfo {author} {\bibfnamefont {B.-j.}\ \bibnamefont
  {Jeon}}\ and\ \bibinfo {author} {\bibfnamefont {M.}~\bibnamefont
  {Muthukumar}},\ }\href {\doibase 10.1021/mz500404e} {\bibfield  {journal}
  {\bibinfo  {journal} {ACS Macro Lett.}\ }\textbf {\bibinfo {volume} {3}},\
  \bibinfo {pages} {911} (\bibinfo {year} {2014}{\natexlab{b}})}\BibitemShut
  {NoStop}%
\bibitem [{\citenamefont {Brun}\ \emph {et~al.}(2008)\citenamefont {Brun},
  \citenamefont {Pastoriza-Gallego}, \citenamefont {Oukhaled}, \citenamefont
  {MathÃ©}, \citenamefont {Bacri}, \citenamefont {Auvray},\ and\
  \citenamefont {Pelta}}]{Brun2008}%
  \BibitemOpen
  \bibfield  {author} {\bibinfo {author} {\bibfnamefont {L.}~\bibnamefont
  {Brun}}, \bibinfo {author} {\bibfnamefont {M.}~\bibnamefont
  {Pastoriza-Gallego}}, \bibinfo {author} {\bibfnamefont {G.}~\bibnamefont
  {Oukhaled}}, \bibinfo {author} {\bibfnamefont {J.}~\bibnamefont {MathÃ©}},
  \bibinfo {author} {\bibfnamefont {L.}~\bibnamefont {Bacri}}, \bibinfo
  {author} {\bibfnamefont {L.}~\bibnamefont {Auvray}}, \ and\ \bibinfo {author}
  {\bibfnamefont {J.}~\bibnamefont {Pelta}},\ }\href {\doibase
  10.1103/physrevlett.100.158302} {\bibfield  {journal} {\bibinfo  {journal}
  {Phys. Rev. Lett.}\ }\textbf {\bibinfo {volume} {100}},\ \bibinfo {pages}
  {158302} (\bibinfo {year} {2008})}\BibitemShut {NoStop}%
\bibitem [{\citenamefont {Murphy}\ and\ \citenamefont
  {Muthukumar}(2007)}]{Murphy2007}%
  \BibitemOpen
  \bibfield  {author} {\bibinfo {author} {\bibfnamefont {R.~J.}\ \bibnamefont
  {Murphy}}\ and\ \bibinfo {author} {\bibfnamefont {M.}~\bibnamefont
  {Muthukumar}},\ }\href {\doibase 10.1063/1.2435717} {\bibfield  {journal}
  {\bibinfo  {journal} {J. Chem. Phys.}\ }\textbf {\bibinfo {volume} {126}},\
  \bibinfo {pages} {051101} (\bibinfo {year} {2007})}\BibitemShut {NoStop}%
\bibitem [{\citenamefont {Meller}, \citenamefont {Nivon},\ and\ \citenamefont
  {Branton}(2001)}]{Meller2001}%
  \BibitemOpen
  \bibfield  {author} {\bibinfo {author} {\bibfnamefont {A.}~\bibnamefont
  {Meller}}, \bibinfo {author} {\bibfnamefont {L.}~\bibnamefont {Nivon}}, \
  and\ \bibinfo {author} {\bibfnamefont {D.}~\bibnamefont {Branton}},\ }\href
  {\doibase 10.1103/physrevlett.86.3435} {\bibfield  {journal} {\bibinfo
  {journal} {Phys. Rev. Lett.}\ }\textbf {\bibinfo {volume} {86}},\ \bibinfo
  {pages} {3435} (\bibinfo {year} {2001})}\BibitemShut {NoStop}%
\bibitem [{\citenamefont {Meller}\ \emph {et~al.}(2000)\citenamefont {Meller},
  \citenamefont {Nivon}, \citenamefont {Brandin}, \citenamefont {Golovchenko},\
  and\ \citenamefont {Branton}}]{Meller2000}%
  \BibitemOpen
  \bibfield  {author} {\bibinfo {author} {\bibfnamefont {A.}~\bibnamefont
  {Meller}}, \bibinfo {author} {\bibfnamefont {L.}~\bibnamefont {Nivon}},
  \bibinfo {author} {\bibfnamefont {E.}~\bibnamefont {Brandin}}, \bibinfo
  {author} {\bibfnamefont {J.}~\bibnamefont {Golovchenko}}, \ and\ \bibinfo
  {author} {\bibfnamefont {D.}~\bibnamefont {Branton}},\ }\href {\doibase
  10.1073/pnas.97.3.1079} {\bibfield  {journal} {\bibinfo  {journal} {P. Natl.
  Acad. Sci. U.S.A.}\ }\textbf {\bibinfo {volume} {97}},\ \bibinfo {pages}
  {1079} (\bibinfo {year} {2000})}\BibitemShut {NoStop}%
\bibitem [{\citenamefont {Wanunu}\ \emph {et~al.}(2008)\citenamefont {Wanunu},
  \citenamefont {Sutin}, \citenamefont {McNally}, \citenamefont {Chow},\ and\
  \citenamefont {Meller}}]{Wanunu2008}%
  \BibitemOpen
  \bibfield  {author} {\bibinfo {author} {\bibfnamefont {M.}~\bibnamefont
  {Wanunu}}, \bibinfo {author} {\bibfnamefont {J.}~\bibnamefont {Sutin}},
  \bibinfo {author} {\bibfnamefont {B.}~\bibnamefont {McNally}}, \bibinfo
  {author} {\bibfnamefont {A.}~\bibnamefont {Chow}}, \ and\ \bibinfo {author}
  {\bibfnamefont {A.}~\bibnamefont {Meller}},\ }\href {\doibase
  10.1529/biophysj.108.140475} {\bibfield  {journal} {\bibinfo  {journal}
  {Biophys. J.}\ }\textbf {\bibinfo {volume} {95}},\ \bibinfo {pages} {4716}
  (\bibinfo {year} {2008})}\BibitemShut {NoStop}%
\bibitem [{\citenamefont {Fologea}\ \emph {et~al.}(2005)\citenamefont
  {Fologea}, \citenamefont {Uplinger}, \citenamefont {Thomas}, \citenamefont
  {McNabb},\ and\ \citenamefont {Li}}]{Fologea2005}%
  \BibitemOpen
  \bibfield  {author} {\bibinfo {author} {\bibfnamefont {D.}~\bibnamefont
  {Fologea}}, \bibinfo {author} {\bibfnamefont {J.}~\bibnamefont {Uplinger}},
  \bibinfo {author} {\bibfnamefont {B.}~\bibnamefont {Thomas}}, \bibinfo
  {author} {\bibfnamefont {D.~S.}\ \bibnamefont {McNabb}}, \ and\ \bibinfo
  {author} {\bibfnamefont {J.}~\bibnamefont {Li}},\ }\href {\doibase
  10.1021/nl051063o} {\bibfield  {journal} {\bibinfo  {journal} {Nano Lett.}\
  }\textbf {\bibinfo {volume} {5}},\ \bibinfo {pages} {1734} (\bibinfo {year}
  {2005})}\BibitemShut {NoStop}%
\bibitem [{\citenamefont {de~Haan}\ and\ \citenamefont
  {Slater}(2012)}]{deHaan2012}%
  \BibitemOpen
  \bibfield  {author} {\bibinfo {author} {\bibfnamefont {H.~W.}\ \bibnamefont
  {de~Haan}}\ and\ \bibinfo {author} {\bibfnamefont {G.~W.}\ \bibnamefont
  {Slater}},\ }\href {\doibase 10.1063/1.4711865} {\bibfield  {journal}
  {\bibinfo  {journal} {J. Chem. Phys.}\ }\textbf {\bibinfo {volume} {136}},\
  \bibinfo {pages} {204902} (\bibinfo {year} {2012})}\BibitemShut {NoStop}%
\bibitem [{\citenamefont {Ding}\ \emph {et~al.}(2012)\citenamefont {Ding},
  \citenamefont {Yan}, \citenamefont {Wang}, \citenamefont {Wu},\ and\
  \citenamefont {Wu}}]{Ding2012}%
  \BibitemOpen
  \bibfield  {author} {\bibinfo {author} {\bibfnamefont {K.}~\bibnamefont
  {Ding}}, \bibinfo {author} {\bibfnamefont {Q.}~\bibnamefont {Yan}}, \bibinfo
  {author} {\bibfnamefont {N.}~\bibnamefont {Wang}}, \bibinfo {author}
  {\bibfnamefont {F.}~\bibnamefont {Wu}}, \ and\ \bibinfo {author}
  {\bibfnamefont {Z.}~\bibnamefont {Wu}},\ }\href {\doibase
  10.1051/epjap/2012110290} {\bibfield  {journal} {\bibinfo  {journal} {Eur.
  Phys. J.-Appl. Phys.}\ }\textbf {\bibinfo {volume} {58}},\ \bibinfo {pages}
  {31201} (\bibinfo {year} {2012})}\BibitemShut {NoStop}%
\bibitem [{\citenamefont {Rasmussen}, \citenamefont {Vishnyakov},\ and\
  \citenamefont {Neimark}(2012)}]{Rasmussen2012}%
  \BibitemOpen
  \bibfield  {author} {\bibinfo {author} {\bibfnamefont {C.~J.}\ \bibnamefont
  {Rasmussen}}, \bibinfo {author} {\bibfnamefont {A.}~\bibnamefont
  {Vishnyakov}}, \ and\ \bibinfo {author} {\bibfnamefont {A.~V.}\ \bibnamefont
  {Neimark}},\ }\href {\doibase 10.1063/1.4754632} {\bibfield  {journal}
  {\bibinfo  {journal} {J. Chem. Phys.}\ }\textbf {\bibinfo {volume} {137}},\
  \bibinfo {pages} {144903} (\bibinfo {year} {2012})}\BibitemShut {NoStop}%
\bibitem [{\citenamefont {Maglia}\ \emph {et~al.}(2008)\citenamefont {Maglia},
  \citenamefont {Restrepo}, \citenamefont {Mikhailova},\ and\ \citenamefont
  {Bayley}}]{Maglia2008}%
  \BibitemOpen
  \bibfield  {author} {\bibinfo {author} {\bibfnamefont {G.}~\bibnamefont
  {Maglia}}, \bibinfo {author} {\bibfnamefont {M.~R.}\ \bibnamefont
  {Restrepo}}, \bibinfo {author} {\bibfnamefont {E.}~\bibnamefont
  {Mikhailova}}, \ and\ \bibinfo {author} {\bibfnamefont {H.}~\bibnamefont
  {Bayley}},\ }\href@noop {} {\bibfield  {journal} {\bibinfo  {journal} {P.
  Natl. Acad. Sci. U.S.A.}\ }\textbf {\bibinfo {volume} {105}},\ \bibinfo
  {pages} {19720} (\bibinfo {year} {2008})}\BibitemShut {NoStop}%
\bibitem [{\citenamefont {Anderson}, \citenamefont {Muthukumar},\ and\
  \citenamefont {Meller}(2013)}]{Anderson2013}%
  \BibitemOpen
  \bibfield  {author} {\bibinfo {author} {\bibfnamefont {B.~N.}\ \bibnamefont
  {Anderson}}, \bibinfo {author} {\bibfnamefont {M.}~\bibnamefont
  {Muthukumar}}, \ and\ \bibinfo {author} {\bibfnamefont {A.}~\bibnamefont
  {Meller}},\ }\href {\doibase 10.1021/nn3051677} {\bibfield  {journal}
  {\bibinfo  {journal} {ACS Nano}\ }\textbf {\bibinfo {volume} {7}},\ \bibinfo
  {pages} {1408} (\bibinfo {year} {2013})}\BibitemShut {NoStop}%
\bibitem [{\citenamefont {Kowalczyk}\ \emph {et~al.}(2012)\citenamefont
  {Kowalczyk}, \citenamefont {Wells}, \citenamefont {Aksimentiev},\ and\
  \citenamefont {Dekker}}]{Kowalczyk2012}%
  \BibitemOpen
  \bibfield  {author} {\bibinfo {author} {\bibfnamefont {S.~W.}\ \bibnamefont
  {Kowalczyk}}, \bibinfo {author} {\bibfnamefont {D.~B.}\ \bibnamefont
  {Wells}}, \bibinfo {author} {\bibfnamefont {A.}~\bibnamefont {Aksimentiev}},
  \ and\ \bibinfo {author} {\bibfnamefont {C.}~\bibnamefont {Dekker}},\
  }\href@noop {} {\bibfield  {journal} {\bibinfo  {journal} {Nano Lett.}\
  }\textbf {\bibinfo {volume} {12}},\ \bibinfo {pages} {1038} (\bibinfo {year}
  {2012})}\BibitemShut {NoStop}%
\bibitem [{\citenamefont {Wong}\ and\ \citenamefont
  {Muthukumar}(2007)}]{Wong2007}%
  \BibitemOpen
  \bibfield  {author} {\bibinfo {author} {\bibfnamefont {C.~T.~A.}\
  \bibnamefont {Wong}}\ and\ \bibinfo {author} {\bibfnamefont {M.}~\bibnamefont
  {Muthukumar}},\ }\href {\doibase 10.1063/1.2723088} {\bibfield  {journal}
  {\bibinfo  {journal} {J. Chem. Phys.}\ }\textbf {\bibinfo {volume} {126}},\
  \bibinfo {pages} {164903} (\bibinfo {year} {2007})}\BibitemShut {NoStop}%
\bibitem [{\citenamefont {Doi}\ and\ \citenamefont {Edwards}(1987)}]{Doi1987}%
  \BibitemOpen
  \bibfield  {author} {\bibinfo {author} {\bibfnamefont {M.}~\bibnamefont
  {Doi}}\ and\ \bibinfo {author} {\bibfnamefont {S.~F.}\ \bibnamefont
  {Edwards}},\ }\href@noop {} {\emph {\bibinfo {title} {The theory of polymer
  dynamics}}}\ (\bibinfo  {publisher} {Oxford [Oxfordshire]: Clarendon Press},\
  \bibinfo {year} {1987})\BibitemShut {NoStop}%
\bibitem [{\citenamefont {Muthukumar}(1987)}]{Muthukumar1987}%
  \BibitemOpen
  \bibfield  {author} {\bibinfo {author} {\bibfnamefont {M.}~\bibnamefont
  {Muthukumar}},\ }\href {\doibase 10.1063/1.452763} {\bibfield  {journal}
  {\bibinfo  {journal} {J. Chem. Phys.}\ }\textbf {\bibinfo {volume} {86}},\
  \bibinfo {pages} {7230} (\bibinfo {year} {1987})}\BibitemShut {NoStop}%
\bibitem [{\citenamefont {Beer}, \citenamefont {Schmidt},\ and\ \citenamefont
  {Muthukumar}(1997)}]{Beer1997}%
  \BibitemOpen
  \bibfield  {author} {\bibinfo {author} {\bibfnamefont {M.}~\bibnamefont
  {Beer}}, \bibinfo {author} {\bibfnamefont {M.}~\bibnamefont {Schmidt}}, \
  and\ \bibinfo {author} {\bibfnamefont {M.}~\bibnamefont {Muthukumar}},\
  }\href {\doibase 10.1021/ma9709821} {\bibfield  {journal} {\bibinfo
  {journal} {Macromolecules}\ }\textbf {\bibinfo {volume} {30}},\ \bibinfo
  {pages} {8375} (\bibinfo {year} {1997})}\BibitemShut {NoStop}%
\bibitem [{\citenamefont {Flory}(1969)}]{Flory1969}%
  \BibitemOpen
  \bibfield  {author} {\bibinfo {author} {\bibfnamefont {P.~J.}\ \bibnamefont
  {Flory}},\ }\href@noop {} {\emph {\bibinfo {title} {Statistical Mechanics of
  Chain Molecules}}}\ (\bibinfo  {publisher} {New York: Interscience
  Publishers},\ \bibinfo {year} {1969})\BibitemShut {NoStop}%
\bibitem [{\citenamefont {Yamakawa}(1971)}]{Yamakawa1971}%
  \BibitemOpen
  \bibfield  {author} {\bibinfo {author} {\bibfnamefont {H.}~\bibnamefont
  {Yamakawa}},\ }\href@noop {} {\emph {\bibinfo {title} {Modern theory of
  polymer solutions}}}\ (\bibinfo  {publisher} {New York: Harper \& Row},\
  \bibinfo {year} {1971})\BibitemShut {NoStop}%
\bibitem [{\citenamefont {Nishida}\ \emph {et~al.}(1997)\citenamefont
  {Nishida}, \citenamefont {Urakawa}, \citenamefont {Kaji}, \citenamefont
  {Gabrys},\ and\ \citenamefont {Higgins}}]{Nishida1997}%
  \BibitemOpen
  \bibfield  {author} {\bibinfo {author} {\bibfnamefont {K.}~\bibnamefont
  {Nishida}}, \bibinfo {author} {\bibfnamefont {H.}~\bibnamefont {Urakawa}},
  \bibinfo {author} {\bibfnamefont {K.}~\bibnamefont {Kaji}}, \bibinfo {author}
  {\bibfnamefont {B.}~\bibnamefont {Gabrys}}, \ and\ \bibinfo {author}
  {\bibfnamefont {J.~S.}\ \bibnamefont {Higgins}},\ }\href {\doibase
  10.1016/s0032-3861(97)00243-7} {\bibfield  {journal} {\bibinfo  {journal}
  {Polymer}\ }\textbf {\bibinfo {volume} {38}},\ \bibinfo {pages} {6083}
  (\bibinfo {year} {1997})}\BibitemShut {NoStop}%
\bibitem [{\citenamefont {Chi}, \citenamefont {Wang},\ and\ \citenamefont
  {Jiang}(2013)}]{Chi2013}%
  \BibitemOpen
  \bibfield  {author} {\bibinfo {author} {\bibfnamefont {Q.}~\bibnamefont
  {Chi}}, \bibinfo {author} {\bibfnamefont {G.}~\bibnamefont {Wang}}, \ and\
  \bibinfo {author} {\bibfnamefont {J.}~\bibnamefont {Jiang}},\ }\href
  {\doibase 10.1016/j.physa.2012.09.022} {\bibfield  {journal} {\bibinfo
  {journal} {Physica A}\ }\textbf {\bibinfo {volume} {392}},\ \bibinfo {pages}
  {1072} (\bibinfo {year} {2013})}\BibitemShut {NoStop}%
\bibitem [{\citenamefont {Eisenriegler}(1982)}]{Eisenriegler1982}%
  \BibitemOpen
  \bibfield  {author} {\bibinfo {author} {\bibfnamefont {E.}~\bibnamefont
  {Eisenriegler}},\ }\href {\doibase 10.1063/1.443835} {\bibfield  {journal}
  {\bibinfo  {journal} {J. Chem. Phys.}\ }\textbf {\bibinfo {volume} {77}},\
  \bibinfo {pages} {6296} (\bibinfo {year} {1982})}\BibitemShut {NoStop}%
\bibitem [{\citenamefont {Muthukumar}(2003)}]{Muthukumar2003}%
  \BibitemOpen
  \bibfield  {author} {\bibinfo {author} {\bibfnamefont {M.}~\bibnamefont
  {Muthukumar}},\ }\href {\doibase 10.1063/1.1553753} {\bibfield  {journal}
  {\bibinfo  {journal} {J. Chem. Phys.}\ }\textbf {\bibinfo {volume} {118}},\
  \bibinfo {pages} {5174} (\bibinfo {year} {2003})}\BibitemShut {NoStop}%
\bibitem [{\citenamefont {Sung}\ and\ \citenamefont {Park}(1996)}]{Sung1996}%
  \BibitemOpen
  \bibfield  {author} {\bibinfo {author} {\bibfnamefont {W.}~\bibnamefont
  {Sung}}\ and\ \bibinfo {author} {\bibfnamefont {P.~J.}\ \bibnamefont
  {Park}},\ }\href {\doibase 10.1103/PhysRevLett.77.783} {\bibfield  {journal}
  {\bibinfo  {journal} {Phys. Rev. Lett.}\ }\textbf {\bibinfo {volume} {77}},\
  \bibinfo {pages} {783} (\bibinfo {year} {1996})}\BibitemShut {NoStop}%
\bibitem [{\citenamefont {Polson}\ and\ \citenamefont
  {McCaffrey}(2013)}]{McCaffrey2013}%
  \BibitemOpen
  \bibfield  {author} {\bibinfo {author} {\bibfnamefont {J.~M.}\ \bibnamefont
  {Polson}}\ and\ \bibinfo {author} {\bibfnamefont {A.~C.~M.}\ \bibnamefont
  {McCaffrey}},\ }\href@noop {} {\bibfield  {journal} {\bibinfo  {journal} {J.
  Chem. Phys.}\ }\textbf {\bibinfo {volume} {138}},\ \bibinfo {eid} {174902}
  (\bibinfo {year} {2013})}\BibitemShut {NoStop}%
\bibitem [{\citenamefont {Polson}\ and\ \citenamefont {Dunn}(2014)}]{Dunn2014}%
  \BibitemOpen
  \bibfield  {author} {\bibinfo {author} {\bibfnamefont {J.~M.}\ \bibnamefont
  {Polson}}\ and\ \bibinfo {author} {\bibfnamefont {T.~R.}\ \bibnamefont
  {Dunn}},\ }\href@noop {} {\bibfield  {journal} {\bibinfo  {journal} {J. Chem.
  Phys.}\ }\textbf {\bibinfo {volume} {140}},\ \bibinfo {eid} {184904}
  (\bibinfo {year} {2014})}\BibitemShut {NoStop}%
\bibitem [{\citenamefont {Muthukumar}\ and\ \citenamefont
  {Katkar}(2015)}]{Muthukumar2015}%
  \BibitemOpen
  \bibfield  {author} {\bibinfo {author} {\bibfnamefont {M.}~\bibnamefont
  {Muthukumar}}\ and\ \bibinfo {author} {\bibfnamefont {H.}~\bibnamefont
  {Katkar}},\ }\href {\doibase 10.1016/j.bpj.2014.11.3452} {\bibfield
  {journal} {\bibinfo  {journal} {Biophys. J.}\ }\textbf {\bibinfo {volume}
  {108}},\ \bibinfo {pages} {17} (\bibinfo {year} {2015})}\BibitemShut
  {NoStop}%
\bibitem [{\citenamefont {Metzler}\ and\ \citenamefont
  {Klafter}(2003)}]{Metzler2003}%
  \BibitemOpen
  \bibfield  {author} {\bibinfo {author} {\bibfnamefont {R.}~\bibnamefont
  {Metzler}}\ and\ \bibinfo {author} {\bibfnamefont {J.}~\bibnamefont
  {Klafter}},\ }\href {\doibase 10.1016/s0006-3495(03)74699-2} {\bibfield
  {journal} {\bibinfo  {journal} {Biophys. J.}\ }\textbf {\bibinfo {volume}
  {85}},\ \bibinfo {pages} {2776} (\bibinfo {year} {2003})}\BibitemShut
  {NoStop}%
\bibitem [{\citenamefont {Dubbeldam}\ \emph
  {et~al.}(2007{\natexlab{a}})\citenamefont {Dubbeldam}, \citenamefont
  {Milchev}, \citenamefont {Rostiashvili},\ and\ \citenamefont
  {Vilgis}}]{Dubbeldam2007}%
  \BibitemOpen
  \bibfield  {author} {\bibinfo {author} {\bibfnamefont {J.~L.~A.}\
  \bibnamefont {Dubbeldam}}, \bibinfo {author} {\bibfnamefont {A.}~\bibnamefont
  {Milchev}}, \bibinfo {author} {\bibfnamefont {V.~G.}\ \bibnamefont
  {Rostiashvili}}, \ and\ \bibinfo {author} {\bibfnamefont {T.~A.}\
  \bibnamefont {Vilgis}},\ }\href {\doibase 10.1209/0295-5075/79/18002}
  {\bibfield  {journal} {\bibinfo  {journal} {Europhys. Lett.}\ }\textbf
  {\bibinfo {volume} {79}},\ \bibinfo {pages} {18002} (\bibinfo {year}
  {2007}{\natexlab{a}})}\BibitemShut {NoStop}%
\bibitem [{\citenamefont {Dubbeldam}\ \emph
  {et~al.}(2007{\natexlab{b}})\citenamefont {Dubbeldam}, \citenamefont
  {Milchev}, \citenamefont {Rostiashvili},\ and\ \citenamefont
  {Vilgis}}]{Dubbeldam2007a}%
  \BibitemOpen
  \bibfield  {author} {\bibinfo {author} {\bibfnamefont {J.~L.~A.}\
  \bibnamefont {Dubbeldam}}, \bibinfo {author} {\bibfnamefont {A.}~\bibnamefont
  {Milchev}}, \bibinfo {author} {\bibfnamefont {V.~G.}\ \bibnamefont
  {Rostiashvili}}, \ and\ \bibinfo {author} {\bibfnamefont {T.~A.}\
  \bibnamefont {Vilgis}},\ }\href {\doibase 10.1103/physreve.76.010801}
  {\bibfield  {journal} {\bibinfo  {journal} {Phys. Rev. E}\ }\textbf {\bibinfo
  {volume} {76}},\ \bibinfo {pages} {010801} (\bibinfo {year}
  {2007}{\natexlab{b}})}\BibitemShut {NoStop}%
\bibitem [{\citenamefont {Dubbeldam}\ \emph {et~al.}(2013)\citenamefont
  {Dubbeldam}, \citenamefont {Rostiashvili}, \citenamefont {Milchev},\ and\
  \citenamefont {Vilgis}}]{Dubbeldam2013}%
  \BibitemOpen
  \bibfield  {author} {\bibinfo {author} {\bibfnamefont {J.~L.~A.}\
  \bibnamefont {Dubbeldam}}, \bibinfo {author} {\bibfnamefont {V.~G.}\
  \bibnamefont {Rostiashvili}}, \bibinfo {author} {\bibfnamefont
  {A.}~\bibnamefont {Milchev}}, \ and\ \bibinfo {author} {\bibfnamefont
  {T.~A.}\ \bibnamefont {Vilgis}},\ }\href {\doibase
  10.1103/physreve.87.032147} {\bibfield  {journal} {\bibinfo  {journal} {Phys.
  Rev. E}\ }\textbf {\bibinfo {volume} {87}},\ \bibinfo {pages} {032147}
  (\bibinfo {year} {2013})}\BibitemShut {NoStop}%
\bibitem [{\citenamefont {Sakaue}(2007)}]{Sakaue2007}%
  \BibitemOpen
  \bibfield  {author} {\bibinfo {author} {\bibfnamefont {T.}~\bibnamefont
  {Sakaue}},\ }\href {\doibase 10.1103/PhysRevE.76.021803} {\bibfield
  {journal} {\bibinfo  {journal} {Phys. Rev. E}\ }\textbf {\bibinfo {volume}
  {76}},\ \bibinfo {pages} {021803} (\bibinfo {year} {2007})}\BibitemShut
  {NoStop}%
\bibitem [{\citenamefont {Sakaue}(2010)}]{Sakaue2010}%
  \BibitemOpen
  \bibfield  {author} {\bibinfo {author} {\bibfnamefont {T.}~\bibnamefont
  {Sakaue}},\ }\href {\doibase 10.1103/PhysRevE.81.041808} {\bibfield
  {journal} {\bibinfo  {journal} {Phys. Rev. E}\ }\textbf {\bibinfo {volume}
  {81}},\ \bibinfo {pages} {041808} (\bibinfo {year} {2010})}\BibitemShut
  {NoStop}%
\bibitem [{\citenamefont {Rowghanian}\ and\ \citenamefont
  {Grosberg}(2011)}]{Rowghanian2011}%
  \BibitemOpen
  \bibfield  {author} {\bibinfo {author} {\bibfnamefont {P.}~\bibnamefont
  {Rowghanian}}\ and\ \bibinfo {author} {\bibfnamefont {A.~Y.}\ \bibnamefont
  {Grosberg}},\ }\href {\doibase 10.1021/jp204014r} {\bibfield  {journal}
  {\bibinfo  {journal} {J. Phys. Chem. B}\ }\textbf {\bibinfo {volume} {115}},\
  \bibinfo {pages} {14127} (\bibinfo {year} {2011})}\BibitemShut {NoStop}%
\bibitem [{\citenamefont {Rowghanian}\ and\ \citenamefont
  {Grosberg}(2012)}]{Rowghanian2012}%
  \BibitemOpen
  \bibfield  {author} {\bibinfo {author} {\bibfnamefont {P.}~\bibnamefont
  {Rowghanian}}\ and\ \bibinfo {author} {\bibfnamefont {A.~Y.}\ \bibnamefont
  {Grosberg}},\ }\href {\doibase 10.1103/PhysRevE.86.011803} {\bibfield
  {journal} {\bibinfo  {journal} {Phys. Rev. E}\ }\textbf {\bibinfo {volume}
  {86}},\ \bibinfo {pages} {011803} (\bibinfo {year} {2012})}\BibitemShut
  {NoStop}%
\bibitem [{\citenamefont {Saito}\ and\ \citenamefont
  {Sakaue}(2011)}]{Saito2011}%
  \BibitemOpen
  \bibfield  {author} {\bibinfo {author} {\bibfnamefont {T.}~\bibnamefont
  {Saito}}\ and\ \bibinfo {author} {\bibfnamefont {T.}~\bibnamefont {Sakaue}},\
  }\href {\doibase 10.1140/epje/i2011-11135-3} {\bibfield  {journal} {\bibinfo
  {journal} {Eur. Phys. J. E}\ }\textbf {\bibinfo {volume} {34}},\ \bibinfo
  {pages} {135} (\bibinfo {year} {2011})}\BibitemShut {NoStop}%
\bibitem [{\citenamefont {Saito}\ and\ \citenamefont
  {Sakaue}(2012{\natexlab{b}})}]{Saito2012err}%
  \BibitemOpen
  \bibfield  {author} {\bibinfo {author} {\bibfnamefont {T.}~\bibnamefont
  {Saito}}\ and\ \bibinfo {author} {\bibfnamefont {T.}~\bibnamefont {Sakaue}},\
  }\href {\doibase 10.1140/epje/i2012-12125-7} {\bibfield  {journal} {\bibinfo
  {journal} {Eur. Phys. J. E}\ }\textbf {\bibinfo {volume} {35}},\ \bibinfo
  {pages} {125} (\bibinfo {year} {2012}{\natexlab{b}})}\BibitemShut {NoStop}%
\bibitem [{\citenamefont {Ikonen}\ \emph
  {et~al.}(2012{\natexlab{a}})\citenamefont {Ikonen}, \citenamefont
  {Bhattacharya}, \citenamefont {Ala-Nissila},\ and\ \citenamefont
  {Sung}}]{Ikonen2012}%
  \BibitemOpen
  \bibfield  {author} {\bibinfo {author} {\bibfnamefont {T.}~\bibnamefont
  {Ikonen}}, \bibinfo {author} {\bibfnamefont {A.}~\bibnamefont
  {Bhattacharya}}, \bibinfo {author} {\bibfnamefont {T.}~\bibnamefont
  {Ala-Nissila}}, \ and\ \bibinfo {author} {\bibfnamefont {W.}~\bibnamefont
  {Sung}},\ }\href {\doibase 10.1103/PhysRevE.85.051803} {\bibfield  {journal}
  {\bibinfo  {journal} {Phys. Rev. E}\ }\textbf {\bibinfo {volume} {85}},\
  \bibinfo {pages} {051803} (\bibinfo {year} {2012}{\natexlab{a}})}\BibitemShut
  {NoStop}%
\bibitem [{\citenamefont {Ikonen}\ \emph
  {et~al.}(2012{\natexlab{b}})\citenamefont {Ikonen}, \citenamefont
  {Bhattacharya}, \citenamefont {Ala-Nissila},\ and\ \citenamefont
  {Sung}}]{Ikonen2012a}%
  \BibitemOpen
  \bibfield  {author} {\bibinfo {author} {\bibfnamefont {T.}~\bibnamefont
  {Ikonen}}, \bibinfo {author} {\bibfnamefont {A.}~\bibnamefont
  {Bhattacharya}}, \bibinfo {author} {\bibfnamefont {T.}~\bibnamefont
  {Ala-Nissila}}, \ and\ \bibinfo {author} {\bibfnamefont {W.}~\bibnamefont
  {Sung}},\ }\href {\doibase 10.1063/1.4742188} {\bibfield  {journal} {\bibinfo
   {journal} {J. Chem. Phys.}\ }\textbf {\bibinfo {volume} {137}},\ \bibinfo
  {pages} {085101} (\bibinfo {year} {2012}{\natexlab{b}})}\BibitemShut
  {NoStop}%
\bibitem [{\citenamefont {Ikonen}\ \emph {et~al.}(2013)\citenamefont {Ikonen},
  \citenamefont {Bhattacharya}, \citenamefont {Ala-Nissila},\ and\
  \citenamefont {Sung}}]{Ikonen2013}%
  \BibitemOpen
  \bibfield  {author} {\bibinfo {author} {\bibfnamefont {T.}~\bibnamefont
  {Ikonen}}, \bibinfo {author} {\bibfnamefont {A.}~\bibnamefont
  {Bhattacharya}}, \bibinfo {author} {\bibfnamefont {T.}~\bibnamefont
  {Ala-Nissila}}, \ and\ \bibinfo {author} {\bibfnamefont {W.}~\bibnamefont
  {Sung}},\ }\href {\doibase 10.1209/0295-5075/103/38001} {\bibfield  {journal}
  {\bibinfo  {journal} {Europhys. Lett.}\ }\textbf {\bibinfo {volume} {103}},\
  \bibinfo {pages} {38001} (\bibinfo {year} {2013})}\BibitemShut {NoStop}%
\bibitem [{\citenamefont {Sean}, \citenamefont {de~Haan},\ and\ \citenamefont
  {Slater}(2015)}]{Sean2015}%
  \BibitemOpen
  \bibfield  {author} {\bibinfo {author} {\bibfnamefont {D.}~\bibnamefont
  {Sean}}, \bibinfo {author} {\bibfnamefont {H.~W.}\ \bibnamefont {de~Haan}}, \
  and\ \bibinfo {author} {\bibfnamefont {G.~W.}\ \bibnamefont {Slater}},\
  }\href {\doibase 10.1002/elps.201400418} {\bibfield  {journal} {\bibinfo
  {journal} {Electrophoresis}\ }\textbf {\bibinfo {volume} {36}},\ \bibinfo
  {pages} {682} (\bibinfo {year} {2015})}\BibitemShut {NoStop}%
\bibitem [{\citenamefont {Plimpton}(1995)}]{Plimpton1995}%
  \BibitemOpen
  \bibfield  {author} {\bibinfo {author} {\bibfnamefont {S.}~\bibnamefont
  {Plimpton}},\ }\href {\doibase 10.1006/jcph.1995.1039} {\bibfield  {journal}
  {\bibinfo  {journal} {J. Comput. Phys.}\ }\textbf {\bibinfo {volume} {117}},\
  \bibinfo {pages} {1} (\bibinfo {year} {1995})}\BibitemShut {NoStop}%
\bibitem [{\citenamefont {Malevanets}\ and\ \citenamefont
  {Kapral}(1999)}]{Malevanets1999}%
  \BibitemOpen
  \bibfield  {author} {\bibinfo {author} {\bibfnamefont {A.}~\bibnamefont
  {Malevanets}}\ and\ \bibinfo {author} {\bibfnamefont {R.}~\bibnamefont
  {Kapral}},\ }\href {\doibase 10.1063/1.478857} {\bibfield  {journal}
  {\bibinfo  {journal} {J. Chem. Phys.}\ }\textbf {\bibinfo {volume} {110}},\
  \bibinfo {pages} {8605} (\bibinfo {year} {1999})}\BibitemShut {NoStop}%
\bibitem [{\citenamefont {Malevanets}\ and\ \citenamefont
  {Yeomans}(2000)}]{Malevanets2000}%
  \BibitemOpen
  \bibfield  {author} {\bibinfo {author} {\bibfnamefont {A.}~\bibnamefont
  {Malevanets}}\ and\ \bibinfo {author} {\bibfnamefont {J.~M.}\ \bibnamefont
  {Yeomans}},\ }\href {\doibase 10.1209/epl/i2000-00428-0} {\bibfield
  {journal} {\bibinfo  {journal} {Europhys. Lett.}\ }\textbf {\bibinfo {volume}
  {52}},\ \bibinfo {pages} {231} (\bibinfo {year} {2000})}\BibitemShut
  {NoStop}%
\bibitem [{\citenamefont {Ihle}\ and\ \citenamefont {Kroll}(2001)}]{Ihle2001}%
  \BibitemOpen
  \bibfield  {author} {\bibinfo {author} {\bibfnamefont {T.}~\bibnamefont
  {Ihle}}\ and\ \bibinfo {author} {\bibfnamefont {D.~M.}\ \bibnamefont
  {Kroll}},\ }\href {\doibase 10.1103/physreve.63.020201} {\bibfield  {journal}
  {\bibinfo  {journal} {Phys. Rev. E}\ }\textbf {\bibinfo {volume} {63}},\
  \bibinfo {pages} {020201} (\bibinfo {year} {2001})}\BibitemShut {NoStop}%
\bibitem [{\citenamefont {de~Haan}, \citenamefont {Sean},\ and\ \citenamefont
  {Slater}(2015)}]{deHaan2015}%
  \BibitemOpen
  \bibfield  {author} {\bibinfo {author} {\bibfnamefont {H.~W.}\ \bibnamefont
  {de~Haan}}, \bibinfo {author} {\bibfnamefont {D.}~\bibnamefont {Sean}}, \
  and\ \bibinfo {author} {\bibfnamefont {G.~W.}\ \bibnamefont {Slater}},\
  }\href@noop {} {\bibfield  {journal} {\bibinfo  {journal} {Phys. Rev. E}\
  }\textbf {\bibinfo {volume} {91}},\ \bibinfo {pages} {022601} (\bibinfo
  {year} {2015})}\BibitemShut {NoStop}%
\bibitem [{\citenamefont {Adhikari}\ and\ \citenamefont
  {Bhattacharya}(2013)}]{Adhikari2013}%
  \BibitemOpen
  \bibfield  {author} {\bibinfo {author} {\bibfnamefont {R.}~\bibnamefont
  {Adhikari}}\ and\ \bibinfo {author} {\bibfnamefont {A.}~\bibnamefont
  {Bhattacharya}},\ }\href {\doibase 10.1063/1.4807002} {\bibfield  {journal}
  {\bibinfo  {journal} {J. Chem. Phys.}\ }\textbf {\bibinfo {volume} {138}},\
  \bibinfo {pages} {204909} (\bibinfo {year} {2013})}\BibitemShut {NoStop}%
\bibitem [{\citenamefont {Cherayil}(1992)}]{Cherayil1992}%
  \BibitemOpen
  \bibfield  {author} {\bibinfo {author} {\bibfnamefont {B.~J.}\ \bibnamefont
  {Cherayil}},\ }\href {\doibase 10.1063/1.463147} {\bibfield  {journal}
  {\bibinfo  {journal} {J. Chem. Phys.}\ }\textbf {\bibinfo {volume} {97}},\
  \bibinfo {pages} {2090} (\bibinfo {year} {1992})}\BibitemShut {NoStop}%
\bibitem [{\citenamefont {Hoogerheide}, \citenamefont {Albertorio},\ and\
  \citenamefont {Golovchenko}(2013)}]{Hoogerheide2013}%
  \BibitemOpen
  \bibfield  {author} {\bibinfo {author} {\bibfnamefont {D.~P.}\ \bibnamefont
  {Hoogerheide}}, \bibinfo {author} {\bibfnamefont {F.}~\bibnamefont
  {Albertorio}}, \ and\ \bibinfo {author} {\bibfnamefont {J.~A.}\ \bibnamefont
  {Golovchenko}},\ }\href {\doibase 10.1103/physrevlett.111.248301} {\bibfield
  {journal} {\bibinfo  {journal} {Phys. Rev. Lett.}\ }\textbf {\bibinfo
  {volume} {111}},\ \bibinfo {pages} {248301} (\bibinfo {year}
  {2013})}\BibitemShut {NoStop}%
\end{thebibliography}
%

\end{document}